\documentclass[prl,numerical,preprint,showkeys]{revtex4-1}

\usepackage{epsfig,amsmath,amssymb,txfonts,hyperref}
\usepackage{pdfpages} 

\makeatletter 
\newcommand*{\ifdraft}[2]{\preprintsty@sw{#1}{#2}}
\AtBeginDocument{\let\LS@rot\@undefined}
\makeatother

\newcommand*{\et}[0]{\textit{et~al.}}

\DeclareMathOperator{\arginf}{arg\ inf}
\DeclareMathOperator{\Tr}{Tr}

\newcommand*{\ox}[0]{\otimes}

\newcommand*{\VECIT}[1]{\mathbf{#1}}
\newcommand*{\VECITS}[1]{\boldsymbol{#1}}  
\newcommand*{\kB}[0]{k_{\text{B}}}

\newcommand*{\xv}[0]{\VECIT{x}}

\newcommand*{\phiv}[0]{\VECITS{\phi}}

\newcommand*{\yv}[0]{\VECIT{y}}

\newcommand*{\bv}[0]{\VECIT{b}}

\newcommand*{\fv}[0]{\VECIT{f}}

\newcommand*{\flux}[1]{\VECIT{J}^{#1}}

\newcommand*{\Onsager}[1]{\underline L^{(#1)}}
\newcommand*{\Onsagert}[1]{\Onsager{\text{#1}}}

\newcommand*{\etav}[0]{\VECITS{\eta}}

\newcommand*{\dxv}[0]{\VECITS{\delta}\VECIT{x}}

\newlength{\wholefigwidth}
\newlength{\smallfigwidth}
\newlength{\halfsmallfigwidth}
\newlength{\figwidth}
\newcommand{\Fig}[1]{Fig.~\ref{fig:#1}}

\newcommand{\Eqn}[1]{Eqn.~\ref{eqn:#1}}

\newcommand{\rcite}[1]{Ref.~\onlinecite{#1}}

\begin{document}

\title{A variational principle for mass transport}

\author{Dallas R. Trinkle}
\email{dtrinkle@illinois.edu}
\affiliation{Department of Materials Science and Engineering, University of Illinois, Urbana-Champaign, Illinois 61801, USA}

\date{\today}
\begin{abstract}
  A variation principle for mass transport in solids is derived that recasts transport coefficients as minima of local thermodynamic average quantities. The result is independent of diffusion mechanism, and applies to amorphous and crystalline systems. This unifies different computational approaches for diffusion, and provides a framework for the creation of new approximation methods with error estimation. It gives a different physical interpretation of the Green function. Finally, the variational principle quantifies the accuracy of competing approaches for a nontrivial diffusion problem.
\end{abstract}

\maketitle

Mass transport in solids is the fundamental kinetic process controlling both the evolution of materials towards equilibrium and a variety of material properties\cite{Balluffi-Kinetics}. Diffusion of atoms dictates everything from the stability of amorphous materials at finite temperature, the design of nanoscaled semiconductor devices, the processing of structural metals including steels and superalloys, the performance of batteries and fuel cells, to the degradation of materials due to corrosion or even irradiation. Since Einstein\cite{Einstein1905}, diffusion has been understood as mesoscale motion arising from many individual atomic displacements, with significant effort over the last century to experimentally measure and model theoretically\cite{Allnatt1993,Mehrer2007}. In the last forty years, computation has played an increasingly important role, with different competing approximation methods developing, combined with increasingly accurate methods to compute transition state energies for atomic processes in transport\cite{Janotti2004,Mantina2009,Garnier2014c}. However, while we have increasing accuracy in predicting atomic scale mechanisms, we lack a clear methodology to compare accuracy of theoretical models that derive mesoscale transport coefficients.

The modern macroscale description of mass transport comes from Onsager's work on nonequilibrium thermodynamics\cite{Onsager1931}, where atomic fluxes $\flux{}$ are linearly proportion to small driving forces. A general driving force is the gradient of chemical potential of species $\alpha$. Then, the Onsager transport coefficients are second-rank tensors $\Onsager{\alpha\beta}$ that relate steady state fluxes in species $\alpha$
\begin{equation}
  \flux{\alpha} = -\sum_\beta \Onsager{\alpha\beta}\nabla\mu^\beta
  \label{eqn:OnsagerFlux}
\end{equation}
are steady-state fluxes in response to perturbatively small driving forces in all other chemical species $\nabla\mu^\beta$. These transport coefficients can also be derived from a thermodynamic extremal principle\cite{Onsager1945,Fischer2014} for maximum entropy production, making the Onsager matrix symmetric and positive-definite.

A brief, albeit incomplete list of methods to compute transport coefficients from atomic mechanisms 
include stochastic methods like kinetic Monte Carlo\cite{Murch1984, Belova2000, Belova2001, Belova2003a, Belova2003b}, master-equation methods like the self-consistent mean-field method\cite{Nastar2000, Nastar2005} and kinetic mean-field approximations\cite{Belashchenko1998, Vaks2014, Vaks2016}, path probability methods for irreversible thermodynamics\cite{Kikuchi1966, Sato1983, Sato1985}, Green function methods\cite{Montroll1965, Koiwa1983, TrinkleElastodiffusivity2016, TrinkleOnsager2017}, and Ritz variational methods\cite{Gortel2004, ZaluskaKotur2007, ZaluskaKotur2014}. The different approaches all have different computational and theoretical complexity, rely on different approximations which may or may not be controlled. However, the relationships between different approximations is not always clear, and it is difficult to determine which of two different calculations is more accurate, short of comparison to experimental results. In what follows, we derive a general expression for the mass transport coefficients in a solid system, and then cast this non-local form into an equivalent minimization problem over \textit{thermodynamic averages of local quantities}: a variational principle for mass transport, with a simple physical interpretation. We show that different computational approaches can be derived and compared with this principle, while also providing a framework for the development of new types of approximations for diffusion. We conclude with a quantitative comparison for a random alloy on a square lattice.

Consider a system with chemical species%
\footnote{The species can include vacancies as an independent chemical species.}
$\alpha=$ A, B, \ldots, with discrete microstates $\{\chi\}$, and transitions between states. For each state $\chi$ and species $\alpha$, $N^\alpha_\chi$ of that species are at positions $\{\xv^\alpha_{\chi i}: i=1\ldots N^\alpha_\chi\}$. Note that the $\xv^\alpha_{\chi i}$ are themselves functions of the state $\chi$. If each state has an energy $E_\chi$, then in the grand canonical ensemble, the equilibrium probability of occupying a given microstate for chemical potentials $\mu^\alpha$ at temperature T is
\begin{equation}
  P^0_\chi:=P^0_\chi(T, \mu^\text{A}, \ldots) = \exp\left[\frac{1}{\kB T} \left(\Phi_0 + \sum_\alpha \mu^\alpha N^\alpha_\chi - E_\chi\right)\right]
  \label{eqn:probability}
\end{equation}
where $\Phi_0$ is a normalization constant---the grand potential---such that $\sum_\chi P^0_\chi = 1$. If the chemical potentials were spatially \textit{inhomogeneous}, then the term corresponding to the sum over chemistry would be $\sum_\alpha\sum_i \mu^\alpha(\xv^\alpha_{\chi i})$. We assume that our system can achieve equilibrium through a Markovian process, with transition rates $W(\chi\to\chi')\ge 0$; then, by detailed balance, $P^0_\chi W(\chi\to\chi') = P^0_{\chi'}W(\chi'\to\chi)$. If all nonzero rates conserve chemical species, then the rates $W(\chi\to\chi')$ are independent of the chemical potentials, and can only depend on the initial and final states and temperature. The master equation for the evolution of a time dependent probability $P_\chi(t)$ is
\begin{equation}
  \frac{dP_\chi(t)}{dt} = \sum_{\chi'} P_{\chi'}(t) W_{\chi'\chi}
  \label{eqn:master}
\end{equation}
and we introduce the shorthand matrix form
\begin{equation}
  W_{\chi'\chi}=
  \begin{cases}
    W(\chi'\to\chi)&:\chi\ne\chi'\\
    -\sum_{\chi'} W(\chi\to\chi')&:\chi=\chi'
  \end{cases}
  \label{ratematrix}
\end{equation}
We identify steady state solutions---which may not be equilibrium solutions---as distributions where the right-hand side is zero for every $\chi$; we are interested in  steady-state solutions that maintain infinitesimal gradients in chemical potentials, for which we will compute fluxes.

What follows is a generalization of results derived previously for a lattice gas model\cite{TrinkleOnsager2017}; details are available in the supplemental material%
\footnote{See Supplemental Material for detailed derivations of transport coefficients, their invariance, equivalence of kinetic Monte Carlo to the variational principle, and different approximation methods based on the variational principle.}.
Consider a steady-state probability distribution $P^\text{ss}_\chi:=P^\text{ss}_\chi(T, \mu^\text{A}, \ldots, \nabla\mu^\text{A}, \ldots)$ in the presence of infinitesimally small chemical potential gradient vectors $\nabla\mu^\alpha$. This steady-state probability distribution can have time-independent fluxes $\flux{\alpha}$ corresponding to mass transport. For any (non-zero rate) transition $\chi\to\chi'$, we define the mass transport vector for each species $\alpha$ as $\dxv^\alpha_{\chi\chi'} := \sum_i \xv^{\alpha}_{\chi'i} - \xv^\alpha_{\chi i}$. This is the net change in positions for all atoms of species $\alpha$, as $N^\alpha_\chi=N^\alpha_{\chi'}$ when $W(\chi\to\chi')\ne0$. Then, the flux is
\begin{equation}
  \flux{\alpha} = V_0^{-1}\sum_{\chi\chi'} P^\text{ss}_\chi W_{\chi\chi'}\dxv^\alpha_{\chi\chi'}
  \label{eqn:flux}
\end{equation}
for total system volume $V_0$. We make the ansatz that the steady-state probability distribution for infinitesimal gradients
\begin{equation}
  P^\text{ss}_\chi = P^0_\chi\left[1 + \frac{\delta\Phi_0}{\kB T} + \frac{1}{\kB T}\sum_\alpha\nabla\mu^\alpha\cdot\left(\etav^\alpha_\chi+\sum_{i=1}^{N^\alpha_\chi} \xv^\alpha_{\chi i}\right)\right]
  \label{eqn:steadystate}
\end{equation}
up to first order in $\nabla\mu^\alpha$, where $\delta\Phi_0$ is a change in the normalization relative to the equilibrium distribution, and introducing the \textit{relaxation vectors} $\etav^\alpha_\chi$ that are to-be-determined for each state $\chi$. If we substitute \Eqn{steadystate} into \Eqn{master}, set $dP^\text{ss}_\chi/dt = 0$, apply detailed balance, divide out by $P^0_\chi$, and require that it hold for arbitrary $\nabla\mu^\alpha$, we find
\begin{equation}
  \sum_{\chi'} W(\chi\to\chi')\dxv^\alpha_{\chi\chi'} = -\sum_{\chi'} W(\chi\to\chi')\left(\etav^\alpha_{\chi'}-\etav^\alpha_\chi\right).
  \label{eqn:balance}
\end{equation}
We define the left-hand side as the velocity vector $\bv^\alpha_\chi:=\sum_{\chi'} W_{\chi\chi'}\dxv^\alpha_{\chi\chi'}$, so that \Eqn{balance} becomes
\begin{equation}
  \bv^\alpha_\chi = -\sum_{\chi'} W_{\chi\chi'}\etav^\alpha_{\chi'}.
  \label{eqn:balance-simple}
\end{equation}
for the steady-state ansatz solution to be time invariant. Then, the transport coefficients $\Onsager{\alpha\beta}$ can be found by substituting the steady-state solution into \Eqn{flux}, while explicitly symmetrizing the summation (rewriting as $\frac12 \sum_{\chi\chi'} + \sum_{\chi'\chi}$), which gives
\begin{equation}
  \Onsager{\alpha\beta} = \frac{1}{\kB TV_0}\Bigg\langle
  \frac12\sum_{\chi'}W_{\chi\chi'}\dxv^\alpha_{\chi\chi'}\otimes\dxv^\beta_{\chi\chi'} - \bv^\alpha_\chi\otimes\etav^\beta_\chi\Bigg\rangle_\chi
  \label{eqn:Onsager}
\end{equation}
where the two terms are the ``uncorrelated'' and ``correlated'' contributions to diffusivity\cite{Allnatt1993, Allnatt1987}, and the average is the shorthand for $\sum_\chi P^0_\chi$.

While \Eqn{Onsager} has the form of a simple thermal average, the primary complication is the solution of \Eqn{balance-simple}, which requires the pseudoinversion of the singular rate matrix $W_{\chi\chi'}$ over the entire state space; this is the Green function $G_{\chi\chi'}:=W^+_{\chi\chi'}$. While the rate matrix is local---as there are only a finite number of final states $\chi'$ to transition from any state $\chi$---the Green function is known to be non-local, and difficult to compute in general. However, the governing equation for the relaxation vectors $\etav^\alpha_\chi$ can be recast instead in a variational form by taking advantage of an \textit{invariance} in \Eqn{Onsager}.

First, the separation of \Eqn{Onsager} into correlated and uncorrelated terms is arbitrary\cite{Allnatt1987,Masi1989}. We introduce changes to the positions of atoms in a state while leaving the rate matrix unchanged: Let $\yv^\alpha_\chi$ be the sum of all displacements of atoms of species $\alpha$ in state $\chi$. We can, without loss of generality%
\footnote{This is accomplished by subtracting a constant value from each $\yv^\alpha_\chi$ to ensure that the sums vanish. This constant leaves the mass transport vectors unchanged.},
consider only cases where $\sum_\chi \yv^\alpha_\chi = 0$. Then, the $\yv^\alpha_\chi$ change the displacement, velocity, and relaxation vectors
\begin{equation*}
  \begin{split}
  &\widetilde\dxv^\alpha_{\chi\chi'} = \dxv^\alpha_{\chi\chi'} + \yv^\alpha_{\chi'}-\yv^\alpha_\chi,
  \qquad \widetilde\bv^\alpha_\chi = \bv^\alpha_\chi + \sum_{\chi'} W_{\chi\chi'}\yv^\alpha_{\chi'},\\
  &\widetilde\etav^\alpha_\chi =
  -\sum_{\chi'} G_{\chi\chi'}\widetilde\bv^\alpha_{\chi'} = 
  \etav^\alpha_\chi - \sum_{\chi'\chi''} G_{\chi\chi'}W_{\chi'\chi''}\yv^\alpha_{\chi''} = \etav^\alpha_\chi - \yv^\alpha_\chi\\
  \end{split}
\end{equation*}
as $G$ is the pseudoinverse of $W$, and $\yv^\alpha_\chi$ is orthogonal to the right null space of $W$. Then, the Onsager coefficients are
\begin{widetext}
  \ifdraft{\footnotesize}{}
  \begin{equation*}
    \begin{split}
  &\kB TV_0\widetilde{\underline L}^{\alpha\beta}
  = \Bigg\langle  \frac12\sum_{\chi'}W_{\chi\chi'}\Big(\dxv^\alpha_{\chi\chi'}+\yv^\alpha_{\chi'}-\yv^\alpha_\chi\Big)\otimes\Big(\dxv^\beta_{\chi\chi'}+\yv^\beta_{\chi'}-\yv^\beta_\chi\Big) - \Big(\bv^\alpha_\chi+\sum_{\chi'}W_{\chi\chi'}\yv^\alpha_{\chi'}\Big)\otimes\Big(\etav^\beta_\chi -\yv^\beta_\chi \Big)\Bigg\rangle_\chi\\
  &\quad = \Bigg\langle  \frac12\sum_{\chi'}W_{\chi\chi'}\dxv^\alpha_{\chi\chi'}\otimes\dxv^\beta_{\chi\chi'} \Bigg\rangle_\chi
  + \frac12\sum_{\chi\chi'} P^0_\chi W_{\chi\chi'}
  \Big(\yv^\alpha_{\chi'}-\yv^\alpha_\chi\Big)\otimes\Big(\yv^\beta_{\chi'}-\yv^\beta_\chi\Big)
  + \frac12\sum_{\chi\chi'} P^0_\chi W_{\chi\chi'}
\dxv^\alpha_{\chi\chi'}\otimes\Big(\yv^\beta_{\chi'}-\yv^\beta_\chi\Big)\\
  &\qquad+ \frac12\sum_{\chi\chi'} P^0_\chi W_{\chi\chi'}
\Big(\yv^\alpha_{\chi'}-\yv^\alpha_\chi\Big)\otimes\dxv^\beta_{\chi\chi'}
  - \left\langle\bv^\alpha_\chi\otimes\etav^\beta_\chi\right\rangle_\chi
  + \sum_{\chi}P^0_\chi\bv^\alpha_\chi\otimes\yv^\beta_\chi
  - \sum_{\chi\chi'}P^0_\chi W_{\chi\chi'}\yv^\alpha_{\chi'}\otimes\etav^\beta_\chi
  + \sum_{\chi\chi'}P^0_\chi W_{\chi\chi'}\yv^\alpha_{\chi'}\otimes\yv^\beta_\chi\\
  &\quad=   \kB TV_0\Onsager{\alpha\beta} 
  - \sum_{\chi\chi'} P^0_\chi W_{\chi\chi'}\yv^\alpha_\chi\otimes\yv^\beta_{\chi'}
  - \left\langle\bv^\alpha_\chi\otimes\yv^\beta_\chi\right\rangle_\chi
  -\left\langle\yv^\alpha_\chi\otimes\bv^\beta_\chi\right\rangle_\chi
  + \left\langle\bv^\alpha_\chi\otimes\yv^\beta_\chi\right\rangle_\chi
  + \left\langle \yv^\alpha_{\chi'}\otimes\bv^\beta_\chi\right\rangle_\chi
  + \sum_{\chi\chi'}P^0_\chi W_{\chi\chi'}\yv^\alpha_{\chi'}\otimes\yv^\beta_\chi
  = \kB TV_0\Onsager{\alpha\beta}\\
    \end{split}
  \label{eqn:invariance}
  \end{equation*}
\end{widetext}
This requires detailed balance $P^0_\chi W_{\chi\chi'}=P^0_{\chi'}W_{\chi'\chi}$ and the sum rule $\sum_{\chi'} W_{\chi\chi'}=0$. Hence, the transport coefficients are invariant under arbitrary displacements, while the ``uncorrelated'' and ``correlated'' terms themselves change.

We can exploit this invariance by noting that, for $\alpha=\beta$, the uncorrelated contribution is positive definite and the correlated contribution is negative definite, as $W_{\chi\chi'}$ and $G_{\chi\chi'}$ are negative definite matrices. Thus, the maximum value of the correlated contribution is zero, which corresponds with the \textit{minimal value} of the \textit{uncorrelated} contribution, and so the equation for the transport coefficients can be rewritten as
\begin{equation}
  \Onsager{\alpha\alpha} = \frac{1}{2\kB TV_0}\inf_{\yv^\alpha_\chi}\Bigg\langle \sum_{\chi'}W_{\chi\chi'}\widetilde\dxv^\alpha_{\chi\chi'}\otimes\widetilde\dxv^\alpha_{\chi\chi'}\Bigg\rangle_\chi,
  \label{eqn:variational}
\end{equation}
which is a \textit{variational principle for mass transport} involving only thermodynamic averages of local quantities.
Here, the infimum of the tensor corresponds to the tensor with the smallest trace%
\footnote{This follows as the correlated contribution is a symmetric negative definite matrix, and the largest value possible is 0, which is achieved when the trace is similarly maximized.}.
The values of $\yv^\alpha_\chi$ that minimize \Eqn{variational} are found by making the generalized force from the gradient of $\ell^\alpha:=\kB TV_0\Tr\Onsager{\alpha\alpha}_\text{uncorr}=\frac12 \left\langle\sum_{\chi'} W_{\chi\chi'}(\widetilde\dxv^\alpha_{\chi\chi'})^2\right\rangle_\chi$,
\begin{equation}
  \begin{split}
    \fv^\alpha_\chi &:=-\frac{\partial\ell^\alpha}{\partial\yv^\alpha_\chi} = -\frac12\frac{\partial}{\partial\yv^\alpha_\chi} \sum_{\chi'\chi''} P^0_{\chi'}W_{\chi'\chi''}\Big(\dxv^\alpha_{\chi'\chi''}+\yv^\alpha_{\chi''}-\yv^\alpha_{\chi'}\Big)^2\\
    &= -2P^0_\chi\sum_{\chi'} W_{\chi\chi'}\Big(\dxv^\alpha_{\chi\chi'}+\yv^\alpha_{\chi'}-\yv^\alpha_\chi\Big)
    = -2P^0_\chi\widetilde\bv^\alpha_\chi
  \end{split}
\end{equation}
equal to zero; this is satisfied when $\yv^\alpha_\chi = \etav^\alpha_\chi$. Moreover, the arguments $\yv^\alpha_\chi$ that minimize $\ell^\alpha$ can then be used to compute the off-diagonal contributions,
\begin{equation}
  \Onsager{\alpha\beta} = \frac{1}{2\kB TV_0}\Bigg\langle \sum_{\chi'}W_{\chi\chi'}\widetilde\dxv^\alpha_{\chi\chi'}\otimes\widetilde\dxv^\beta_{\chi\chi'}\Bigg\rangle_{\chi}\;\Bigg|_{\yv^\alpha_\chi = \arginf \ell^\alpha}
\end{equation}
This is similar to the Varadhan-Spohn variational form\cite{Spohn1991}, which Arita~\et\ note is a powerful, albeit abstract result that is difficult to apply in practice, involving ``cylinder'' functions\cite{Arita2017}. Note that \Eqn{variational} is simpler than the alternate Ritz variational form, as there is no normalization of a eigenvector required\cite{Gortel2004, ZaluskaKotur2007, ZaluskaKotur2014}.

This variational principle for mass transport has multiple consequences. First, it unifies multiple approaches for the computation of mass transport coefficients, including kinetic Monte Carlo, Green function methods, and self-consistent mean-field theory. Moreover, it provides a direct way to compare the accuracy of different methods: outside of the convergence of stochastic sampling errors, once a mass transport method is recast in a variational form, the minimal value of the diagonal transport coefficients is necessarily closer to the true value. It also gives a simple physical explanation for the correlation contributions in mass transport: the $\etav^\alpha_\chi$ values are displacements that map a correlated random walk into an equivalent uncorrelated random walk with identical transport coefficients. Finally, it provides a framework for the construction of new algorithms for the computation of mass transport that requires the minimization of a thermal average; as it is based on minimization, different approximations for $\yv^\alpha_\chi$ can be simultaneously introduced, while the process of minimization finds the optimal solution.

In the case of a linear expansion for the relaxation vectors, the variational principle for mass transport provides a simple general expression for diffusivity. Let $\{\phiv^\alpha_{\chi,n}\}$ be a set of basis vectors so that we expand $\yv^\alpha_\chi = \sum_n \phiv^\alpha_{\chi,n}\theta^\alpha_n$ with coefficients $\theta^\alpha_n$. The supplemental material Sec.~S4 shows the most general solution; here, we include the solution for the case where the basis functions are chemistry- and direction-independent: $\phiv^\alpha_{\chi,ni}=\hat e_i \phi_{\chi,n}$ for a Cartesian orthonormal basis $\hat e_1, \hat e_2, \hat e_3$. Then, the coefficients that minimize \Eqn{variational} can be found by solving $\sum_m \overline W_{nm}\theta^\alpha_{mi} = \overline\bv^\alpha_n\cdot\hat e_i$ where,
\begin{equation}
    \overline W_{nm} := \Bigg\langle \sum_{\chi'} W_{\chi\chi'}\phi_{\chi,n}\phi^\alpha_{\chi',m}\Bigg\rangle_{\chi},\quad
    \overline\bv^\alpha_n := \left\langle \phi_{\chi,n}\bv^\alpha_\chi\right\rangle_{\chi}.
  \label{eqn:averagerate}
\end{equation}
We can take the pseudoinverse of $\overline G := (\overline W)^+$, and then the transport coefficients are (c.f. Eqn.~S24)
\begin{multline}
  \Onsager{\alpha\beta}_\text{LBAM} = \frac1{2\kB TV_0}\Bigg\langle \sum_{\chi'} W_{\chi\chi'} \dxv^\alpha_{\chi\chi'}\ox\dxv^\beta_{\chi\chi'}\Bigg\rangle_\chi
  \ifdraft{}{\\}
  +\frac1{\kB TV_0}\sum_{nm} \left\langle\phi_{\chi,n}\bv^\alpha_{\chi}\right\rangle_\chi\ox \overline G_{nm}\left\langle \phi_{\chi,m} \bv^\beta_\chi\right\rangle_{\chi}
\end{multline}
where the diagonal transport coefficients $\Onsager{\alpha\alpha}_\text{LBAM}$ are guaranteed to be an upper bound on the true coefficients, achieving equality when the basis $\{\phi_{\chi,n}\}$ spans $\etav^\alpha_\chi$.

We can now express existing computational approaches as attempts to solve the variational problem. For kinetic Monte Carlo\cite{Murch1984, Belova2000, Belova2001, Belova2003a, Belova2003b}, each trajectory represents a single sample in the average, while the increasing length of a trajectory attempts to converge the relaxation vectors corresponding to that single starting state. In Sec.~S3, the equivalence of kinetic Monte Carlo to the variation method is shown; moreover, the use of a finite length trajectory is variational: assuming perfect sampling of initial states, and with perfect sampling of trajectories of a finite length, the transport coefficients will be greater than the true transport coefficients. If one uses accelerated KMC methods\cite{Novotny1995, Puchala2010, Nandipati2010, Chatterjee2010, Fichthorn2013}, superbasins---a finite collection of states with fast internal transitions but slow escapes---are effectively collapsed onto a single \textit{position}, which is an approximation to the relaxation vector $\etav^\alpha_\chi$. For vacancy-mediated diffusion, the dilute Green function\cite{TrinkleElastodiffusivity2016, TrinkleOnsager2017} and matrix methodology\cite{Montroll1965, Koiwa1983} work in a restricted state space $\{\chi\}$ where only one solute and vacancy are present, and then effectively construct a full basis in that state space. Finally, self-consistent mean-field\cite{Nastar2000, Nastar2005} and kinetic mean-field\cite{Belashchenko1998, Vaks2014, Vaks2016} work with a cluster expansion of chemistry- and direction-independent basis functions $\{\phi_{\chi,n}\}$ that are products of site occupancies for different chemistries. It should be noted that these latter two methods derive their solution for the parameters $\theta_n$ using a ladder of $n$-body correlation functions on which they invoke ``closure approximations'' for higher order correlation functions; in a variational framework, such closure approximations become unnecessary. Finally, when methods are framed in variational terms, we can quantitatively compare accuracy by identifying which method gives the smallest diagonal elements $\Onsager{\alpha\alpha}$, and also estimate remaining error through the average residual bias, $\langle (\widetilde\bv^\alpha_\chi)^2/(-W_{\chi\chi})\rangle_\chi$ in Eqn.~S27, or its ratio with $\langle (\bv^\alpha_\chi)^2/(-W_{\chi\chi})\rangle_\chi$.

In addition to providing a common frame for existing computational methods for mass transport, we now have a new framework to develop and test new approximations, including those that are more appropriate for amorphous systems that lack crystalline order but still possess well-defined microstates. A simple example is the basis function choice $\phiv^\alpha_{\chi}=\bv^\alpha_\chi$; in Sec.~S5 of the supplemental material, a closed-form approximation for transport coefficients is provided in Eqn.~S30. This approximation involves inverting a matrix that has the same dimensionality as the number of independent chemical species; however, it only captures local correlations. We can also take the dilute Green function methodology for vacancy-mediated transport into finite solute concentrations by using the basis functions $\phi_{\chi,\beta\xv}$ that are equal to the occupancy (0 or 1) by chemistry $\beta$ of a site at a vector $\xv$ relative to a vacancy in state $\chi$. This approximation exactly reproduces the dilute solute limit by being equivalent to an infinite range two-body-only version of the Green function.

For a quantitative comparison of these new approximations, we consider a random binary alloy on a square lattice with a single vacancy. In this model, there is no binding energy between any species, and the jump rate for the vacancy only depends on the chemistry of the species it is exchanging: either $\nu_\text{A}$ (``solvent'' exchange) or $\nu_\text{B}$ (``solute'' exchange). We take $\nu_\text{A}=1$, and consider three cases: $\nu_\text{B}=1$ (tracer), $\nu_\text{B}=4$ (``fast'' diffuser), and $\nu_\text{B}=0$ (frozen solute). This system has nontrivial behavior, including a percolation limit\cite{Moleko1989, Allnatt2016} for $\nu_\text{B}=0$ where the diffusivity of solvent is 0 for $c_\text{B}<1$. To compute the transport coefficients, we use: (1) kinetic Monte Carlo on a $64\times64$ periodic grid, generating 256 samples of trajectories run for 4096 vacancy jumps each; the transport coefficients are computed 32 separate times to get a mean and stochastic error estimate. (2) A two-body Green function approximation (c.f. Sec.~S6), which has the analytic solution (c.f. Eqn.~S38),
\begin{equation}
  \begin{split}
    \Onsagert{AA}_\text{GF} &= \mathbf{1}c_\text{v} a_0^2\left[c_\text{A}\nu_\text{A} -
      \frac{c_\text{A}c_\text{B}\nu^2_\text{A}}{\nu_\text{A} + \nu_\text{B} + \frac{2f-1}{1-f}(c_\text{A}\nu_\text{A}+c_\text{B}\nu_\text{B})}\right]\\
    \Onsagert{BB}_\text{GF} &= \mathbf{1}c_\text{v} a_0^2\left[c_\text{B}\nu_\text{B} -
      \frac{c_\text{A}c_\text{B}\nu^2_\text{B}}{\nu_\text{A} + \nu_\text{B} + \frac{2f-1}{1-f}(c_\text{A}\nu_\text{A}+c_\text{B}\nu_\text{B})}\right]\\
  \end{split}
  \label{eqn:SCGF}
\end{equation}
where $f=(\pi-1)^{-1}\approx 0.467$ is the dilute tracer correlation coefficient for a square lattice. (3) A bias basis approximation, which has the same transport coefficients as \Eqn{SCGF} with the approximation $f=1-2/(z+1)=0.6$. (4) A self-consistent mean-field approach taking into account clusters of all orders within two jumps: $\pm\hat x$, $\pm\hat y$, $\pm\hat x\pm\hat y$, $\pm2\hat x$, and $\pm2\hat y$. Finally, for $\nu_\text{B}=0$ we use a direct solution for vacancy diffusivity with 256 configurations of a $256\times256$ periodic cell, and compute a residual bias correction (RBC) for the Green function results.

\begin{figure}[ht]
  \includegraphics[width=0.5\figwidth]{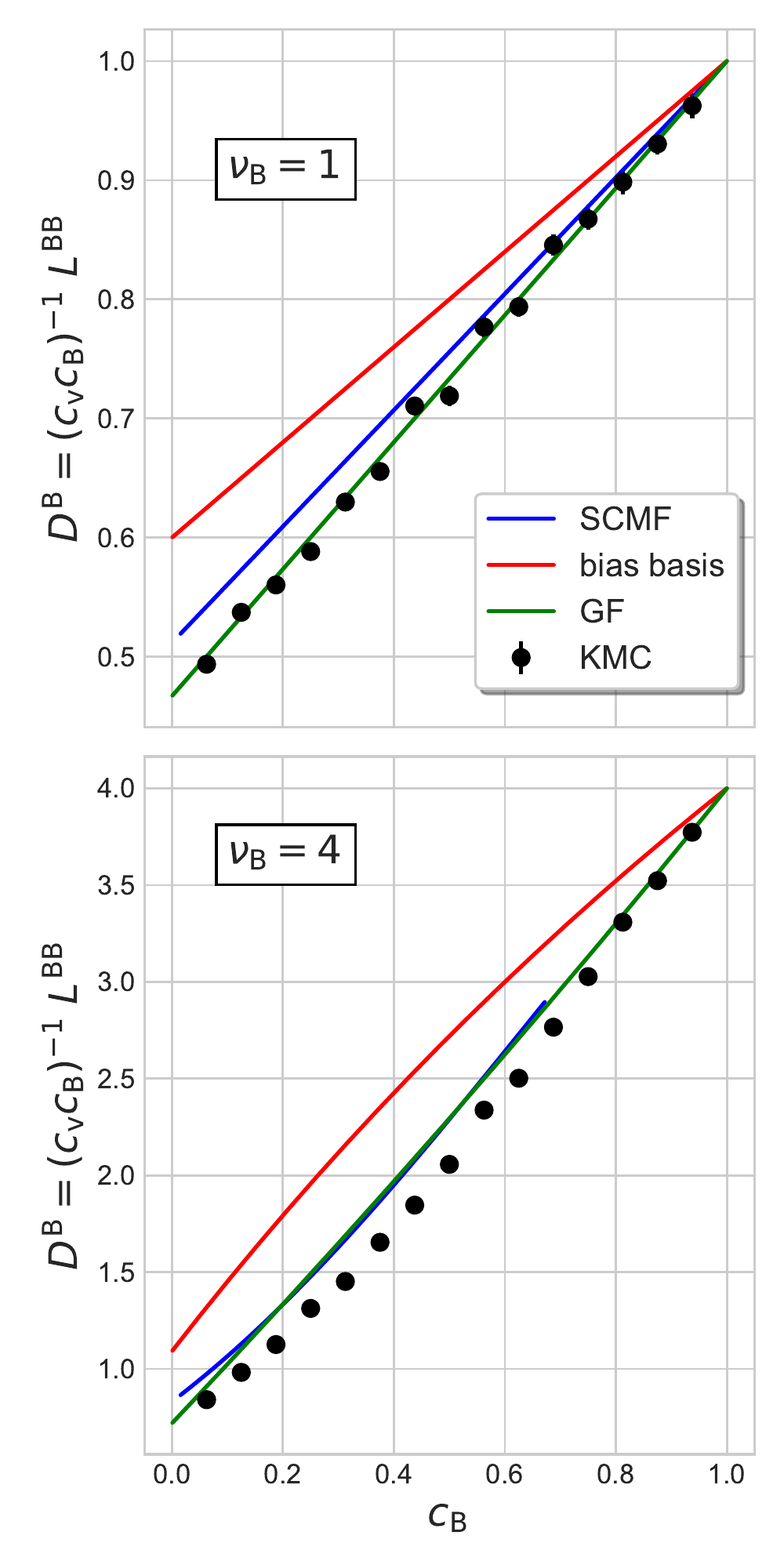}
  \includegraphics[width=0.5\figwidth]{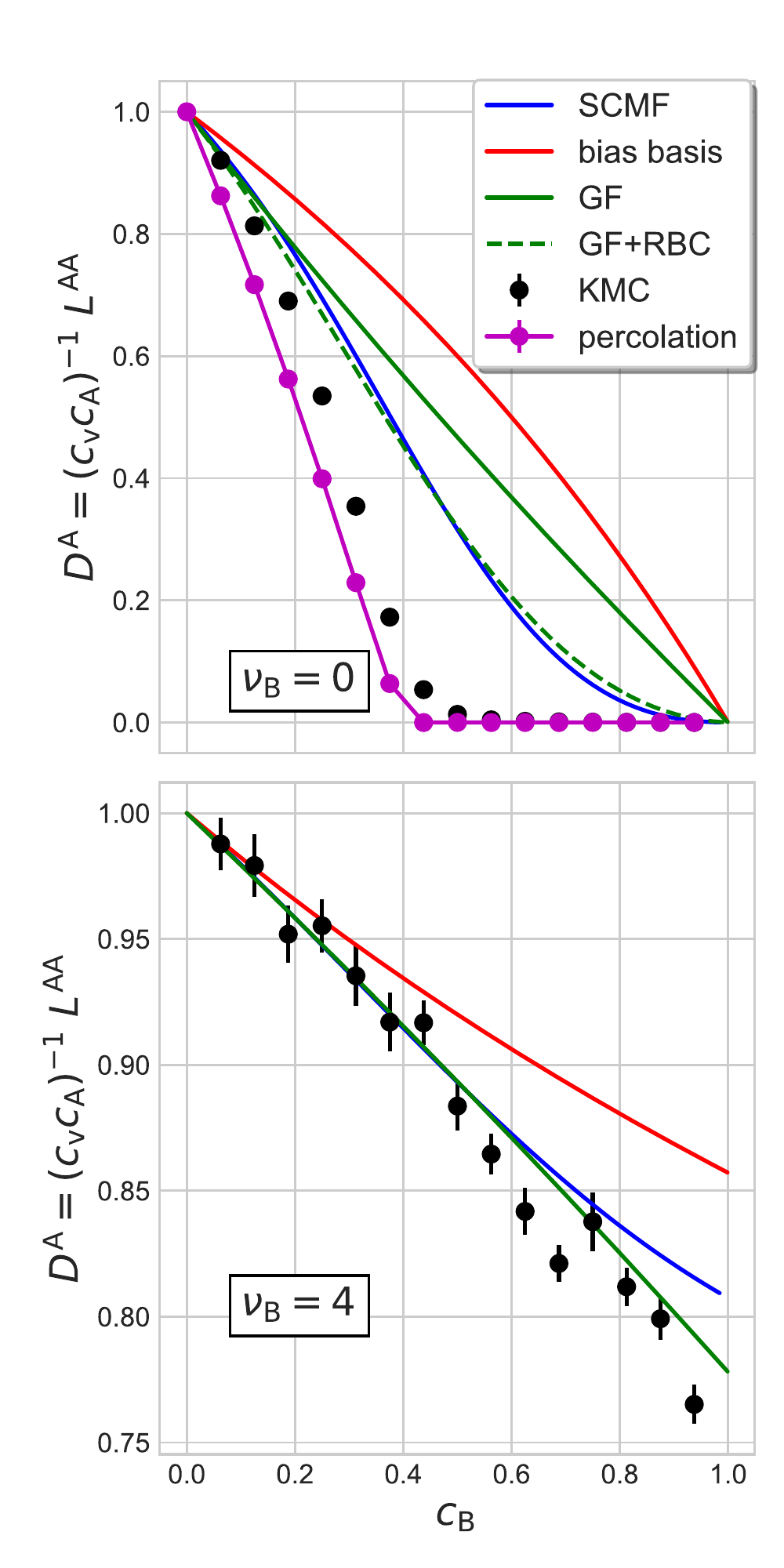}
\caption{Diffusivity of ``solute'' atom B and ``solvent'' A in a random alloy on a square lattice, scaled by vacancy concentration. The vacancy exchange rate with atom A is 1, while we consider different relative rates for vacancy-B exchange and different solute concentrations from 0 to 1. The largest deviation is for the case where $\nu_\text{B}=0$, which requires longer range correlation to capture the percolation limit near $c_\text{B}\approx 0.5$; the ``percolation'' simulations are averages of 256 direct calculations on a $256\times256$ periodic cell.}
\label{fig:diffusivity}
\end{figure}

\Fig{diffusivity} shows the different accuracy for this binary system. The Green function approach captures the dilute A and B limits for the tracer and fast diffuser examples, and is the most accurate of the three approaches. The largest difference is seen for the percolation case $\nu_\text{B}=0$, where both the Green function and self-consistent mean-field methods are good approximations for $c_\text{B}\lesssim 0.2$, but begin to break down as we approach the percolation limit. In this case, solutes are creating islands where a vacancy is trapped and unable to diffuse over long distances; inside such an island, the relaxation vectors should map all ``trapped'' states onto the same position, producing no contribution to the diffusivity. We also see the direct simulations produce lower, more accurate, diffusivity. The size of these islands gets smaller as $c_\text{B}$ increases, and only the self-consistent mean-field method---and only at large concentrations of solute---is able to reproduce the behavior seen by kinetic Monte Carlo. This suggests the need to go beyond the two-body basis for the Green function approach, or combining local multisite basis functions with long-range basis functions, or perhaps new approximation methods all together. One example such approach is the RBC, where following a linear basis approximation method, the residual bias vectors serve as basis vectors for a correction to the diffusivity; in the case of $\nu_\text{B}=0$, we derive an analytic expression (c.f. Sec.~S7, Eqn.~S45) that has similar error to the SCMF result.

With a variational formulation of transport coefficients, we can develop new approximate methods for modeling diffusion in solids, including amorphous materials. If linear approximations are used, then basis functions provide a projection of the state space into a subspace while the variational principle provides a lower bound on transport coefficients. The selection of basis functions can be guided by physical insight, and systematic improvement is always possible. It is also possible to construct \textit{nonlinear} approximations to the relaxation vectors $\yv^\alpha_\chi$ which might require fewer parameters to describe; still, a variational principle permits relative comparisons of different methods, and a lower bound on the result. While the fundamental insight for the variational formulation came from the invariance in \Eqn{Onsager}, it can be derived as an thermodynamic extremum principle where the positions of atoms are ``free'' variables, connecting to Onsager's original work.

\begin{acknowledgments}
  The author thanks Pascal Bellon and Maylise Nastar for helpful conversations and suggestions for the manuscript.
  This research was supported in part by the U.S. Department of Energy, Office of Basic Energy Sciences, Division of Materials Sciences and Engineering under Award \#DE-FG02-05ER46217, through the Frederick Seitz Materials Research Laboratory, in part by the Office of Naval Research grant N000141210752, and in part by the National Science Foundation Award 1411106. Part of this research was performed while the author was visiting the Institute for Pure and Applied Mathematics (IPAM) at UCLA, which is supported by the National Science Foundation (NSF). The python notebook for the random alloy results in \Fig{diffusivity} is available online at \rcite{OnsagerCalc}, which makes use of the algebraic multigrid solver \textsc{PyAMG}\cite{BeOlSc2011}.
\end{acknowledgments}


%

\newpage


\section*{Supplementary material}

# A variational principle for mass transport: Supplemental material

Dallas R. Trinkle*

*Department of Materials Science and Engineering,*

*University of Illinois, Urbana-Champaign, Illinois 61801, USA*

(Dated: May 28, 2018)

## Abstract

Supplemental material: (1) detailed derivation of transport coefficients $\underline{L}^{(\alpha\beta)}$; (2) detailed derivation of invariance of transport coefficients under arbitrary displacements; (3) equivalence of kinetic Monte Carlo calculation of transport coefficients with variational result; (4) detailed derivation of linear basis transport coefficient approximation; (5) detailed derivation of bias basis and scaled bias basis approximation; (6) averaged Green function approximation for a cubic binary random alloy with two exchange rates; (7) the residual bias correction.

---

* dtrinkle@illinois.edu

In the derivations that follow, we will take advantage of detailed balance for the equilibrium probability distribution, which provides a symmetry relation

$$P^0_\chi W_{\chi\chi'} = P^0_{\chi'} W_{\chi'\chi}, \quad \forall \chi, \chi' \tag{S1}$$

and an antisymmetry relation

$$P^0_\chi W_{\chi\chi'} \delta \mathbf{x}^\alpha_{\chi\chi'} = -P^0_{\chi'} W_{\chi'\chi} \delta \mathbf{x}^\alpha_{\chi'\chi}, \quad \forall \alpha, \chi, \chi' \tag{S2}$$

A further consequence is the symmetry relation also applies for the pseudoinverse of the rate matrix, the Green function $G := W^+$

$$P^0_\chi G_{\chi\chi'} = P^0_{\chi'} G_{\chi'\chi}, \quad \forall \chi, \chi' \tag{S3}$$

## S1. TRANSPORT COEFFICIENT DERIVATION

To derive the Onsager transport coefficients for infinitesimal chemical potential gradients, the flux equation must be evaluated with the steady-state probability distribution;

$$\mathbf{J}^\alpha = V_0^{-1} \sum_{\chi\chi'} P^{ss}_\chi W_{\chi\chi'} \delta \mathbf{x}^\alpha_{\chi\chi'} \tag{S4}$$

where

$$P^{ss}_\chi = P^0_\chi \left[ 1 + \frac{\delta \Phi_0}{k_B T} + \frac{1}{k_B T} \sum_\alpha \nabla \mu^\alpha \cdot \left( \boldsymbol{\eta}^\alpha_\chi + \sum_{i=1}^{N^\alpha_\chi} \mathbf{x}^\alpha_{\chi i} \right) \right]. \tag{S5}$$

This distribution has to satisfy the master equation, $dP^{ss}_\chi/dt = \sum_{\chi'} P^{ss}_{\chi'} W_{\chi'\chi} = 0$, or

$$0 = \frac{1}{k_B T} \sum_{\chi'} P^0_{\chi'} W_{\chi'\chi} \left[ k_B T + \delta \Phi_0 + \sum_\alpha \nabla \mu^\alpha \cdot \left( \boldsymbol{\eta}^\alpha_{\chi'} + \sum_{i=1}^{N^\alpha_{\chi'}} \mathbf{x}^\alpha_{\chi'i} \right) \right]$$

$$0 = \frac{P^0_\chi}{k_B T} \sum_\alpha \left[ \sum_{\chi'} W_{\chi\chi'} \boldsymbol{\eta}^\alpha_{\chi'} + \sum_{\chi'} W_{\chi\chi'} \sum_{i=1}^{N^\alpha_{\chi'}} \mathbf{x}^\alpha_{\chi'i} \right] \cdot \nabla \mu^\alpha \tag{S6}$$

$$-\frac{P^0_\chi}{k_B T} \sum_\alpha \left[ \sum_{\chi'} W_{\chi\chi'} \boldsymbol{\eta}^\alpha_{\chi'} \right] \cdot \nabla \mu^\alpha = \frac{P^0_\chi}{k_B T} \mathbf{b}^\alpha_\chi \cdot \nabla \mu^\alpha$$

which simplifies to $-\sum_{\chi'} W_{\chi\chi'} \boldsymbol{\eta}^\alpha_{\chi'} = \mathbf{b}^\alpha_\chi$.

The transport coefficients are most easily derived by rewriting a "symmetrized" version of the flux equation and applying the symmetry and antisymmetry relations Eqn. S1 and Eqn. S2,

$$\begin{aligned}
\mathbf{J}^\alpha &= \frac{1}{2V_0} \sum_{\chi\chi'} \left[ P^{ss}_\chi W_{\chi\chi'} \delta\mathbf{x}^\alpha_{\chi\chi'} + P^{ss}_{\chi'} W_{\chi'\chi} \delta\mathbf{x}^\alpha_{\chi'\chi} \right] \\
&= \frac{1}{2k_B T V_0} \sum_{\chi\chi'} \left[ \left( P^0_\chi W_{\chi\chi'} \delta\mathbf{x}^\alpha_{\chi\chi'} + P^0_{\chi'} W_{\chi'\chi} \delta\mathbf{x}^\alpha_{\chi'\chi} \right)(k_B T + \delta\Phi_0) \right. \\
&\quad \left. + P^0_\chi W_{\chi\chi'} \delta\mathbf{x}^\alpha_{\chi\chi'} \sum_\beta \nabla\mu^\beta \cdot \left( \boldsymbol{\eta}^\beta_\chi + \sum_{i=1}^{N^\beta_\chi} \mathbf{x}^\beta_{\chi i} \right) + P^0_{\chi'} W_{\chi'\chi} \delta\mathbf{x}^\alpha_{\chi'\chi} \sum_\beta \nabla\mu^\beta \cdot \left( \boldsymbol{\eta}^\beta_{\chi'} + \sum_{i=1}^{N^\beta_{\chi'}} \mathbf{x}^\beta_{\chi' i} \right) \right] \\
&= \frac{1}{2k_B T V_0} \sum_\beta \sum_{\chi\chi'} P^0_\chi W_{\chi\chi'} \left[ \delta\mathbf{x}^\alpha_{\chi\chi'} \otimes \left( \boldsymbol{\eta}^\beta_\chi - \boldsymbol{\eta}^\beta_{\chi'} \right) + \delta\mathbf{x}^\alpha_{\chi\chi'} \otimes \left( \sum_{i=1}^{N^\beta_\chi} \mathbf{x}^\beta_{\chi i} - \mathbf{x}^\beta_{\chi' i} \right) \right] \cdot \nabla\mu^\beta \\
&= -\frac{1}{2k_B T V_0} \sum_\beta \left[ \sum_{\chi\chi'} P^0_\chi W_{\chi\chi'} \delta\mathbf{x}^\alpha_{\chi\chi'} \otimes \delta\mathbf{x}^\beta_{\chi\chi'} - \sum_\chi P^0_\chi \left( \sum_{\chi'} W_{\chi\chi'} \delta\mathbf{x}^\alpha_{\chi\chi'} \right) \otimes \boldsymbol{\eta}^\beta_\chi \right. \\
&\quad \left. + \sum_{\chi\chi'} P^0_\chi W_{\chi\chi'} \delta\mathbf{x}^\alpha_{\chi\chi'} \otimes \boldsymbol{\eta}^\beta_{\chi'} \right] \cdot \nabla\mu^\beta \\
&= -\frac{1}{2k_B T V_0} \sum_\beta \left[ \sum_{\chi\chi'} P^0_\chi W_{\chi\chi'} \delta\mathbf{x}^\alpha_{\chi\chi'} \otimes \delta\mathbf{x}^\beta_{\chi\chi'} - 2\sum_\chi P^0_\chi \mathbf{b}^\alpha_\chi \otimes \boldsymbol{\eta}^\beta_\chi \right] \cdot \nabla\mu^\beta \\
&= -\sum_\beta \left[ \frac{1}{2k_B T V_0} \left\langle \sum_{\chi'} W_{\chi\chi'} \delta\mathbf{x}^\alpha_{\chi\chi'} \otimes \delta\mathbf{x}^\beta_{\chi\chi'} \right\rangle_\chi - \frac{1}{k_B T V_0} \left\langle \mathbf{b}^\alpha_\chi \otimes \boldsymbol{\eta}^\beta_\chi \right\rangle_\chi \right] \cdot \nabla\mu^\beta = -\sum_\beta \underline{L}^{(\alpha\beta)} \cdot \nabla\mu^\beta.
\end{aligned}$$
(S7)

which is the result presented in Eqn. 9.

## S2. TRANSPORT COEFFICIENT INVARIANCE DERIVATION

The invariance of the transport coefficients to arbitrary displacements takes advantage of the same symmetry and antisymmetry relations as before. We introduce changes to the positions of atoms in a state by summing all of the displacements $\mathbf{y}^\alpha_\chi$; we impose the restriction $\sum_\chi \mathbf{y}^\alpha_\chi = 0$ to simplify the treatment below. This can be done without loss of generality, for if $\mathbf{y}^\alpha_0 := \sum_\chi \mathbf{y}^\alpha_\chi \neq 0$, the new displacement vectors $\widetilde{\delta\mathbf{x}}^\alpha_{\chi\chi'}$ are unchanged if we use $\mathbf{y}^\alpha_\chi + a\mathbf{y}^\alpha_0$ for any value of $a$. Then, our changed vectors are

$$\widetilde{\delta\mathbf{x}}^\alpha_{\chi\chi'} = \delta\mathbf{x}^\alpha_{\chi\chi'} + \mathbf{y}^\alpha_{\chi'} - \mathbf{y}^\alpha_\chi, \qquad \widetilde{\mathbf{b}}^\alpha_\chi = \mathbf{b}^\alpha_\chi + \sum_{\chi'} W_{\chi\chi'} \mathbf{y}^\alpha_{\chi'}$$

$$\widetilde{\boldsymbol{\eta}}^\alpha_\chi = -\sum_{\chi'} G_{\chi\chi'} \widetilde{\mathbf{b}}^\alpha_{\chi'} = \boldsymbol{\eta}^\alpha_\chi - \sum_{\chi'\chi''} G_{\chi\chi'} W_{\chi'\chi''} \mathbf{y}^\alpha_{\chi''} = \boldsymbol{\eta}^\alpha_\chi - \mathbf{y}^\alpha_\chi$$

(S8)

3

Then,

$$k_\text{B} T V_0 \widetilde{\underline{L}}^{\alpha\beta} = \left\langle \frac{1}{2} \sum_{\chi'} W_{\chi\chi'} \left( \delta \mathbf{x}^{\alpha}_{\chi\chi'} + \mathbf{y}^{\alpha}_{\chi'} - \mathbf{y}^{\alpha}_{\chi} \right) \otimes \left( \delta \mathbf{x}^{\beta}_{\chi\chi'} + \mathbf{y}^{\beta}_{\chi'} - \mathbf{y}^{\beta}_{\chi} \right) - \left( \mathbf{b}^{\alpha}_{\chi} + \sum_{\chi'} W_{\chi\chi'} \mathbf{y}^{\alpha}_{\chi'} \right) \otimes \left( \boldsymbol{\eta}^{\beta}_{\chi} - \mathbf{y}^{\beta}_{\chi} \right) \right\rangle_{\chi}$$

$$= \left\langle \frac{1}{2} \sum_{\chi'} W_{\chi\chi'} \delta \mathbf{x}^{\alpha}_{\chi\chi'} \otimes \delta \mathbf{x}^{\beta}_{\chi\chi'} \right\rangle_{\chi} + \frac{1}{2} \sum_{\chi\chi'} P^0_{\chi} W_{\chi\chi'} \left( \mathbf{y}^{\alpha}_{\chi'} - \mathbf{y}^{\alpha}_{\chi} \right) \otimes \left( \mathbf{y}^{\beta}_{\chi'} - \mathbf{y}^{\beta}_{\chi} \right)$$

$$+ \frac{1}{2} \sum_{\chi\chi'} P^0_{\chi} W_{\chi\chi'} \delta \mathbf{x}^{\alpha}_{\chi\chi'} \otimes \left( \mathbf{y}^{\beta}_{\chi'} - \mathbf{y}^{\beta}_{\chi} \right) + \frac{1}{2} \sum_{\chi\chi'} P^0_{\chi} W_{\chi\chi'} \left( \mathbf{y}^{\alpha}_{\chi'} - \mathbf{y}^{\alpha}_{\chi} \right) \otimes \delta \mathbf{x}^{\beta}_{\chi\chi'}$$

$$- \left\langle \mathbf{b}^{\alpha}_{\chi} \otimes \boldsymbol{\eta}^{\beta}_{\chi} \right\rangle_{\chi} + \sum_{\chi} P^0_{\chi} \mathbf{b}^{\alpha}_{\chi} \otimes \mathbf{y}^{\beta}_{\chi} - \sum_{\chi\chi'} P^0_{\chi} W_{\chi\chi'} \mathbf{y}^{\alpha}_{\chi'} \otimes \boldsymbol{\eta}^{\beta}_{\chi} + \sum_{\chi\chi'} P^0_{\chi} W_{\chi\chi'} \mathbf{y}^{\alpha}_{\chi'} \otimes \mathbf{y}^{\beta}_{\chi}$$

$$= k_\text{B} T V_0 \underline{L}^{(\alpha\beta)} + \frac{1}{2} \sum_{\chi\chi'} P^0_{\chi} W_{\chi\chi'} \mathbf{y}^{\alpha}_{\chi'} \otimes \mathbf{y}^{\beta}_{\chi'} + \frac{1}{2} \sum_{\chi\chi'} P^0_{\chi} W_{\chi\chi'} \mathbf{y}^{\alpha}_{\chi} \otimes \mathbf{y}^{\beta}_{\chi} - \frac{1}{2} \sum_{\chi\chi'} P^0_{\chi} W_{\chi\chi'} \mathbf{y}^{\alpha}_{\chi} \otimes \mathbf{y}^{\beta}_{\chi'}$$

$$- \frac{1}{2} \sum_{\chi\chi'} P^0_{\chi} W_{\chi\chi'} \mathbf{y}^{\alpha}_{\chi'} \otimes \mathbf{y}^{\beta}_{\chi} + \frac{1}{2} \sum_{\chi\chi'} P^0_{\chi} W_{\chi\chi'} \delta \mathbf{x}^{\alpha}_{\chi\chi'} \otimes \mathbf{y}^{\beta}_{\chi'} - \frac{1}{2} \sum_{\chi\chi'} P^0_{\chi} W_{\chi\chi'} \delta \mathbf{x}^{\alpha}_{\chi\chi'} \otimes \mathbf{y}^{\beta}_{\chi}$$

$$+ \frac{1}{2} \sum_{\chi\chi'} P^0_{\chi} W_{\chi\chi'} \mathbf{y}^{\alpha}_{\chi'} \otimes \delta \mathbf{x}^{\beta}_{\chi\chi'} - \frac{1}{2} \sum_{\chi\chi'} P^0_{\chi} W_{\chi\chi'} \mathbf{y}^{\alpha}_{\chi} \otimes \delta \mathbf{x}^{\beta}_{\chi\chi'}$$

$$+ \left\langle \mathbf{b}^{\alpha}_{\chi} \otimes \mathbf{y}^{\beta}_{\chi} \right\rangle_{\chi} - \sum_{\chi\chi'} P^0_{\chi'} W_{\chi'\chi} \mathbf{y}^{\alpha}_{\chi'} \otimes \boldsymbol{\eta}^{\beta}_{\chi} + \sum_{\chi\chi'} P^0_{\chi} W_{\chi\chi'} \mathbf{y}^{\alpha}_{\chi'} \otimes \mathbf{y}^{\beta}_{\chi}$$

$$= k_\text{B} T V_0 \underline{L}^{(\alpha\beta)} - \sum_{\chi\chi'} P^0_{\chi} W_{\chi\chi'} \mathbf{y}^{\alpha}_{\chi} \otimes \mathbf{y}^{\beta}_{\chi'} - \left\langle \mathbf{b}^{\alpha}_{\chi} \otimes \mathbf{y}^{\beta}_{\chi} \right\rangle_{\chi} - \left\langle \mathbf{y}^{\alpha}_{\chi} \otimes \mathbf{b}^{\beta}_{\chi} \right\rangle_{\chi} + \left\langle \mathbf{b}^{\alpha}_{\chi} \otimes \mathbf{y}^{\beta}_{\chi} \right\rangle_{\chi} + \left\langle \mathbf{y}^{\alpha}_{\chi'} \otimes \mathbf{b}^{\beta}_{\chi} \right\rangle_{\chi}$$

$$+ \sum_{\chi\chi'} P^0_{\chi} W_{\chi\chi'} \mathbf{y}^{\alpha}_{\chi'} \otimes \mathbf{y}^{\beta}_{\chi}$$

$$= k_\text{B} T V_0 \underline{L}^{(\alpha\beta)}$$

(S9)

where both the symmetry and antisymmetry relations Eqn. S1 and Eqn. S2 are used to transform terms containing $W\delta \mathbf{x}$ into $\mathbf{b}$ as appropriate. This is the final result in Eqn. .

## S3. EQUIVALENCE OF KINETIC MONTE CARLO APPROXIMATION OF TRANSPORT COEFFICIENTS

To show that kinetic Monte Carlo approximations of the transport coefficients converge to the same result as given by Eqn. 9, we first perform some matrix transformations. The matrix

$$\Gamma_{\chi\chi'} := \begin{cases} -(W_{\chi\chi})^{-1} W_{\chi\chi'} & : \chi \neq \chi' \\ 0 & : \chi = \chi' \end{cases}$$

(S10)

4

gives the probability of selecting the transition from $\chi$ to $\chi'$ in a kinetic Monte Carlo algorithm. We also identify the average residency time of a state $\chi$ as $\tau_\chi := -(W_{\chi\chi})^{-1}$. This allows us to write $W_{\chi\chi'} = W_{\chi\chi}\delta_{\chi\chi'} - W_{\chi\chi}\Gamma_{\chi\chi'}$. Then, the Green function can be rewritten as an infinite series

$$G = \left[\left(\tau^{-1}\right)(\mathbf{1} - \Gamma)\right]^{-1} = (\mathbf{1} - \Gamma)^{-1}\tau = \sum_{n=0}^{\infty}\Gamma^n\tau \tag{S11}$$

We use this form of the Green function to rewrite the relaxation vectors $\boldsymbol{\eta}_\chi^\alpha$,

$$\begin{aligned}
\boldsymbol{\eta}_\chi^\alpha &= -\sum_{\chi'}G_{\chi\chi'}\mathbf{b}_{\chi'}^\alpha = -\sum_{\chi'}G_{\chi\chi'}\sum_{\chi''}W_{\chi'\chi''}\delta\mathbf{x}_{\chi'\chi''}^\alpha \\
&= -\sum_{n=0}^{\infty}\sum_{\chi'}(\Gamma^n)_{\chi\chi'}\sum_{\chi''}\tau_{\chi'}W_{\chi'\chi''}\delta\mathbf{x}_{\chi'\chi''}^\alpha \\
&= \sum_{n=0}^{\infty}\sum_{\chi'\chi''}(\Gamma^n)_{\chi\chi'}\Gamma_{\chi'\chi''}\delta\mathbf{x}_{\chi'\chi''}^\alpha
\end{aligned} \tag{S12}$$

This equation can be interpreted in terms of trajectories. Consider the (finite) sequence of states $\chi_0 \to \chi_1 \to \chi_2 \to \cdots \to \chi_N$, where each successive state is chosen with probability $\Gamma_{\chi\chi'}$. The initial state has a probability $P_{\chi_0}^0$, while any given sequence occurs with the probability

$$\prod_{n=0}^{N-1}\Gamma_{\chi_n,\chi_{n+1}}$$

The $n = 0$ term in Eqn. S12 is the average of displacements from state $\chi$ when it makes one transition; that is, the average displacement from all paths of length 1. The $n = 1$ term is the average displacement from the *second* transition of all paths of length 2; in general, the $n^{\text{th}}$ term is the average displacement of the *last* transition of all paths of length $n + 1$. Thus, the sum of the first $n$ terms is the *total* displacement averaged over all paths of length $n + 1$, and hence, $\boldsymbol{\eta}_\chi^\alpha$ is the average total displacement for all trajectories starting at $\chi$ and run to infinite length.

We can use this trajectory result into Eqn. 9 after we symmetrize

$$\left\langle \mathbf{b}_\chi^\alpha \otimes \boldsymbol{\eta}_\chi^\beta \right\rangle_\chi = \frac{1}{2}\left\langle \mathbf{b}_\chi^\alpha \otimes \boldsymbol{\eta}_\chi^\beta \right\rangle_\chi + \frac{1}{2}\left\langle \boldsymbol{\eta}_\chi^\alpha \otimes \mathbf{b}_\chi^\beta \right\rangle_\chi$$

by using the symmetry relation Eqn. S3. Then,

$$\begin{aligned}
2k_{\text{B}}TV_0\underline{L}^{(\alpha\beta)} &= \sum_{\chi\chi'}P_\chi^0 W_{\chi\chi'}\delta\mathbf{x}_{\chi\chi'}^\alpha \otimes \delta\mathbf{x}_{\chi\chi'}^\beta - \sum_{\chi\chi'}P_\chi^0 W_{\chi\chi'}\delta\mathbf{x}_{\chi\chi'}^\alpha \otimes \boldsymbol{\eta}_\chi^\beta - \sum_{\chi\chi'}P_\chi^0 \boldsymbol{\eta}_\chi^\alpha \otimes W_{\chi\chi'}\delta\mathbf{x}_{\chi\chi'}^\beta \\
&= \sum_{\chi_0\chi_1}P_{\chi_0}^0 W_{\chi_0\chi_1}\delta\mathbf{x}_{\chi_0\chi_1}^\alpha \otimes \delta\mathbf{x}_{\chi_0\chi_1}^\beta + \sum_{\chi_0\chi_1}P_{\chi_0}^0 W_{\chi_0\chi_1}\delta\mathbf{x}_{\chi_0\chi_1}^\alpha \otimes \left(\sum_{n=0}^{\infty}\sum_{\chi'\chi''}(\Gamma^n)_{\chi_1\chi'}\Gamma_{\chi'\chi''}\delta\mathbf{x}_{\chi'\chi''}^\beta\right) \\
&+ \sum_{\chi_0\chi_1}P_{\chi_0}^0 W_{\chi_0\chi_1}\left(\sum_{n=0}^{\infty}\sum_{\chi'\chi''}(\Gamma^n)_{\chi_1\chi'}\Gamma_{\chi'\chi''}\delta\mathbf{x}_{\chi'\chi''}^\alpha\right) \otimes \delta\mathbf{x}_{\chi_0\chi_1}^\beta
\end{aligned} \tag{S13}$$

5

The sign changes in the bias vectors on the second line are accomplished via the antisymmetry relation Eqn. S2.

We finally need to be able to express thermal averages in terms of samples in a trajectory. As we select each successive state using $\Gamma$, we can apply detailed balance to $\Gamma$, as

$$P^0_\chi \tau_\chi^{-1} \Gamma_{\chi\chi'} = P^0_{\chi'} \tau_{\chi'}^{-1} \Gamma_{\chi'\chi}$$

which indicates that a state appears in a trajectory with a probability proportional to $P^0_\chi \tau_\chi^{-1}$. Hence, we can use a trajectory to approximate thermal averages by multiplying by $\tau_\chi$:

$$\langle f_\chi \rangle_\chi \approx \frac{\sum_{n=0}^{N} \tau_{\chi_n} f_{\chi_n}}{\sum_{n=0}^{N} \tau_{\chi_n}} \tag{S14}$$

If we consider a kinetic Monte Carlo algorithm for diffusivity, we (1) select initial states $\chi_0$ with probability $P^0_\chi$, (2) use $\Gamma_{\chi\chi'}$ to select the next state in a trajectory, while accumulating $T := \sum_n \tau_{\chi_n}$ and $\mathbf{x}^\alpha := \sum_n \delta\mathbf{x}^\alpha_{\chi_n\chi_{n+1}}$ for $N$ total transitions; then

$$\underline{L}^{(\alpha\beta)} \approx \underline{L}^{(\alpha\beta)}_{\text{KMC}} := \frac{1}{N_{\text{trajectories}}} \sum_{\text{initial }\chi_0} \frac{\left(\sum_{n=0}^{N-1} \delta\mathbf{x}^\alpha_{\chi_n\chi_{n+1}}\right) \otimes \left(\sum_{m=0}^{N-1} \delta\mathbf{x}^\beta_{\chi_m\chi_{m+1}}\right)}{2k_B T V_0 \sum_{n=0}^{N-1} \tau_{\chi_n}} \tag{S15}$$

which is the usual result based on mean-squared displacements. To connect Eqn. S15 with Eqn. S13, we separate the sums in the numerator into three cases: $n = m$, $n < m$, and $n > m$, and transform from the trajectory approximations back to thermal averages,

$$\begin{aligned}
2k_B T V_0 \underline{L}^{(\alpha\beta)}_{\text{KMC}} &= \frac{1}{N_{\text{trajectories}}} \sum_{\text{initial }\chi_0} \left\{\sum_{n=0}^{N-1} \tau_{\chi_n}\right\}^{-1} \left\{\sum_{n=0}^{N-1} \delta\mathbf{x}^\alpha_{\chi_n\chi_{n+1}} \otimes \delta\mathbf{x}^\beta_{\chi_n\chi_{n+1}}\right.\\
&\left.+ \sum_{n=0}^{N-1} \delta\mathbf{x}^\alpha_{\chi_n\chi_{n+1}} \otimes \left(\sum_{m=n+1}^{N-1} \delta\mathbf{x}^\beta_{\chi_m\chi_{m+1}}\right) + \sum_{m=0}^{N-1} \left(\sum_{n=m+1}^{N-1} \delta\mathbf{x}^\alpha_{\chi_n\chi_{n+1}}\right) \otimes \delta\mathbf{x}^\beta_{\chi_m\chi_{m+1}}\right\}\\
&\approx \sum_{\chi_0} P^0_{\chi_0} \tau_{\chi_0}^{-1} \sum_{\chi_1} \Gamma_{\chi_0\chi_1} \delta\mathbf{x}^\alpha_{\chi_0\chi_1} \otimes \delta\mathbf{x}^\beta_{\chi_0\chi_1}\\
&+ \sum_{\chi_0} P^0_{\chi_0} \tau_{\chi_0}^{-1} \sum_{\chi_1} \Gamma_{\chi_0\chi_1} \delta\mathbf{x}^\alpha_{\chi_0\chi_1} \otimes \delta\mathbf{x}^\beta_{\chi_0\chi_1} \otimes \left(\sum_{n=0}^{\infty} \sum_{\chi'\chi''} (\Gamma^n)_{\chi_1\chi'} \Gamma_{\chi'\chi''} \delta\mathbf{x}^\beta_{\chi'\chi''}\right)\\
&+ \sum_{\chi_0} P^0_{\chi_0} \tau_{\chi_0}^{-1} \left(\sum_{n=0}^{\infty} \sum_{\chi'\chi''} (\Gamma^n)_{\chi_1\chi'} \Gamma_{\chi'\chi''} \delta\mathbf{x}^\beta_{\chi'\chi''}\right) \otimes \sum_{\chi_1} \Gamma_{\chi_0\chi_1} \delta\mathbf{x}^\alpha_{\chi_0\chi_1}\\
&= 2k_B T V_0 \underline{L}^{(\alpha\beta)}
\end{aligned} \tag{S16}$$

6

The primary approximation beyond stochastic sampling is the approximation of an infinite trajectory length with a finite one. This also suggests that the length of the trajectory must be sufficient to converge $\boldsymbol{\eta}_\chi^\alpha$ values, and that the number of trajectories should be sufficient for accurate thermal sampling. It also identifies kinetic Monte Carlo as stochastic minimization technique: increasingly longer pathways help to converge towards the true (minimal) transport coefficients.

The convergence of the trajectory length can be cast in terms of the variational principle as well. If we assume the ability to sample *all* trajectories of a fixed length of steps $N_{\text{KMC}}$, this is equivalent to using an approximate value of the relaxation vector $\boldsymbol{\eta}_\chi^\alpha$. This follows as

$$G - (\Gamma)^{N_{\text{KMC}}} G = \sum_{n=0}^{\infty} \Gamma^n \tau - \sum_{n=N_{\text{KMC}}}^{\infty} \Gamma^n \tau = \sum_{n=0}^{N_{\text{KMC}}-1} \Gamma^n \tau \tag{S17}$$

so the kinetic Monte Carlo with finite length trajectories is computing

$$\boldsymbol{\eta}_{\chi,\text{KMC}}^\alpha = \boldsymbol{\eta}_\chi^\alpha - (\Gamma)^{N_{\text{KMC}}} G \mathbf{b}_\chi^\alpha. \tag{S18}$$

Since $\Gamma$ is positive-definite, while $G$ is negative definite, this means that the kinetic Monte Carlo calculation in Eqn. S13 is adding a positive definite contribution to the Onsager coefficients, which is reduced by increasing $N_{\text{KMC}}$. Hence, finite-length trajectories in kinetic Monte Carlo are also variational solutions, in the limit of sufficient sampling of the initial states and all trajectories of length $N_{\text{KMC}}$.

## S4. LINEAR BASIS APPROXIMATION METHOD FOR TRANSPORT COEFFICIENTS

Let $\{\boldsymbol{\phi}_{\chi,n}^\alpha\}$ be a set of basis vectors so that we expand $\mathbf{y}_\chi^\alpha = \sum_n \boldsymbol{\phi}_{\chi,n}^\alpha \theta_n^\alpha$ with coefficients $\theta_n^\alpha$. This is the most general form of linear basis approximations; one may consider *chemistry- and direction-independent* basis functions $\phi_{\chi,n} \hat{e}_i$ for orthonormal basis vector $\{\hat{e}_1, \hat{e}_2, \hat{e}_3\}$ which have some advantages in their solution, but at the possible expense of requiring a larger basis set. As a trivial example, the chemistry-dependent basis set where $\boldsymbol{\phi}_{\chi,1}^\alpha = \boldsymbol{\eta}_\chi^\alpha$ allows for the exact solution of the transport coefficients with a minimal basis set of size 1. We start with the transport coefficient solutions for the general case, and then specific results for chemistry-independent basis functions.

The uncorrelated transport coefficients as a function of parameters $\theta_n^\alpha$ are

$$k_B T V_0 \underline{L}_{\text{LBAM}}^{(\alpha\beta)}(\theta) = \frac{1}{2}\left\langle \sum_{\chi'} W_{\chi\chi'} \delta \mathbf{x}_{\chi\chi'}^\alpha \otimes \delta \mathbf{x}_{\chi\chi'}^\beta \right\rangle_\chi - \sum_m \left\langle \boldsymbol{\phi}_{\chi,m}^\alpha \otimes \mathbf{b}_\chi^\beta \right\rangle_\chi \theta_m^\alpha - \sum_m \left\langle \mathbf{b}_\chi^\alpha \otimes \boldsymbol{\phi}_{\chi,m}^\beta \right\rangle_\chi \theta_m^\beta$$
$$- \sum_{mm'} \left\langle \sum_{\chi'} W_{\chi\chi'} \boldsymbol{\phi}_{\chi,m}^\alpha \otimes \boldsymbol{\phi}_{\chi,m'}^\beta \right\rangle_\chi \theta_m^\alpha \theta_{m'}^\beta \tag{S19}$$

where we find the optimal solutions by setting the generalized force coefficients to zero,

$$-k_B T V_0 \frac{d}{d\theta_n^\alpha} \operatorname{Tr} \underline{L}_{\text{LBAM}}^{(\alpha\alpha)} = -2 \left\langle \boldsymbol{\phi}_{\chi,n}^\alpha \cdot \mathbf{b}_\chi^\alpha \right\rangle_\chi - 2 \sum_m \left\langle \sum_{\chi'} W_{\chi\chi'} \boldsymbol{\phi}_{\chi,n}^\alpha \cdot \boldsymbol{\phi}_{\chi',m}^\alpha \right\rangle_\chi \theta_m^\alpha = 0 \quad (S20)$$

We can then define

$$\underline{\overline{W}}_{nm}^{\alpha\beta} := \left\langle \sum_{\chi'} W_{\chi\chi'} \boldsymbol{\phi}_{\chi,n}^\alpha \otimes \boldsymbol{\phi}_{\chi',m}^\beta \right\rangle_\chi, \quad \overline{b}_n^\alpha := \left\langle \boldsymbol{\phi}_{\chi,n}^\alpha \cdot \mathbf{b}_\chi^\alpha \right\rangle_\chi$$

$$\overline{W}_{nm}^\alpha := \operatorname{Tr} \underline{\overline{W}}_{nm}^{\alpha\alpha} = \left\langle \sum_{\chi'} W_{\chi\chi'} \boldsymbol{\phi}_{\chi,n}^\alpha \cdot \boldsymbol{\phi}_{\chi',m}^\alpha \right\rangle_\chi, \quad \overline{G}^\alpha := \left(\overline{W}^\alpha\right)^+ \quad (S21)$$

so that the optimal $\theta_n^\alpha$ are $\overline{\eta}_n^\alpha := -\sum_m \overline{G}_{nm}^\alpha \overline{b}_m^\alpha$, and

$$k_B T V_0 \underline{L}_{\text{LBAM}}^{(\alpha\beta)} = \frac{1}{2} \left\langle \sum_{\chi'} W_{\chi\chi'} \delta \mathbf{x}_{\chi\chi'}^\alpha \otimes \delta \mathbf{x}_{\chi\chi'}^\beta \right\rangle_\chi + \sum_{nm} \left\langle \boldsymbol{\phi}_{\chi,n}^\alpha \otimes \mathbf{b}_\chi^\beta \right\rangle_\chi \overline{G}_{nm}^\alpha \left\langle \boldsymbol{\phi}_{\chi,m}^\alpha \cdot \mathbf{b}_\chi^\alpha \right\rangle_\chi$$
$$+ \sum_{nm} \left\langle \mathbf{b}_\chi^\alpha \otimes \boldsymbol{\phi}_{\chi,n}^\beta \right\rangle_\chi \overline{G}_{nm}^\beta \left\langle \boldsymbol{\phi}_{\chi,m}^\beta \cdot \mathbf{b}_\chi^\beta \right\rangle_\chi - \sum_{nmn'm'} \left\langle \mathbf{b}_\chi^\alpha \cdot \boldsymbol{\phi}_{\chi,n}^\alpha \right\rangle_\chi \overline{G}_{nm}^\alpha \underline{\overline{W}}_{mm'}^{\alpha\beta} \overline{G}_{m'n'}^\beta \left\langle \mathbf{b}_\chi^\beta \cdot \boldsymbol{\phi}_{\chi,n'}^\beta \right\rangle_\chi \quad (S22)$$

In the case of chemisty- and direction-independent basis functions $\boldsymbol{\phi}_{\chi,ni} = \phi_{\chi,n} \hat{e}_i$ for orthonormal basis vector $\{\hat{e}_1, \hat{e}_2, \hat{e}_3\}$, the matrices and vectors are simplified as

$$\underline{\overline{W}}_{nimj} := \hat{e}_i \otimes \hat{e}_j \left\langle \sum_{\chi'} W_{\chi\chi'} \phi_{\chi,n} \phi_{\chi',m} \right\rangle_\chi, \quad \overline{\mathbf{b}}_n^\alpha := \left\langle \phi_{\chi,n} \mathbf{b}_\chi^\alpha \right\rangle_\chi$$

$$\overline{W}_{nm} := \sum_i \underline{\overline{W}}_{nimi} = \left\langle \sum_{\chi'} W_{\chi\chi'} \phi_{\chi,n} \phi_{\chi',m} \right\rangle_\chi, \quad \overline{G} := \overline{W}^+ \quad (S23)$$

and then $\overline{\eta}_n^\alpha := -\sum_m \overline{G}_{nm} \overline{\mathbf{b}}_m^\alpha$, and $\overline{G}\,\overline{W}\,\overline{G} = \mathbf{1}$, and the transport coefficients are

$$k_B T V_0 \underline{L}_{\text{LBAM}}^{(\alpha\beta)} = \frac{1}{2} \left\langle \sum_{\chi'} W_{\chi\chi'} \delta \mathbf{x}_{\chi\chi'}^\alpha \otimes \delta \mathbf{x}_{\chi\chi'}^\beta \right\rangle_\chi + \sum_{nm} \left\langle \phi_{\chi,n} \mathbf{b}_\chi^\alpha \right\rangle_\chi \otimes \overline{G}_{nm} \left\langle \phi_{\chi,m} \mathbf{b}_\chi^\beta \right\rangle_\chi \quad (S24)$$

This form most easily matches the dilute-limit Green function method, and the self-consistent mean-field (SCMF) approximation; a finite-ranged cluster expansion basis produces the SCMF method. It can also be used to derive the averaged Green function approximation in Section S6. It is exact when where we expand our basis to all states, where our index $n = \chi'$ so that $\phi_{\chi\chi'} = \delta_{\chi\chi'}$.

We can also compute the average residual bias vector as a quantitative measure of remaining error from the linear basis approximation method. Using $\tau_\chi$ as our shorthand for the inverse escape time $(-W_{\chi\chi})^{-1}$, we can compute

$$R^\alpha := \langle \widetilde{\mathbf{b}}_\chi^\alpha \cdot \widetilde{\mathbf{b}}_\chi^\alpha \tau_\chi \rangle_\chi$$

8

which has the same units as $k_\text{B}TV_0\underline{L}^{(\alpha\alpha)}$. The post-relaxation bias is

$$\widetilde{\mathbf{b}}_\chi^\alpha = \mathbf{b}_\chi^\alpha + \sum_n \sum_{\chi'} W_{\chi\chi'}\boldsymbol{\phi}_{\chi',n}^\alpha \overline{\eta}_n^\alpha \tag{S25}$$

so the residual bias can be simplified by writing $\tau_\chi W_{\chi\chi'} = \Gamma_{\chi\chi'} - \delta_{\chi\chi'}$ and $\tau_\chi^{-1}\Gamma_{\chi\chi'} = W_{\chi\chi'} + \tau_\chi^{-1}\delta_{\chi\chi'}$,

$$\begin{aligned}
R^\alpha &= \langle \mathbf{b}_\chi^\alpha \cdot \mathbf{b}_\chi^\alpha \tau_\chi \rangle_\chi + 2\sum_n \overline{\eta}_n^\alpha \sum_{\chi\chi'} \tau_\chi P_\chi^0 W_{\chi\chi'} \boldsymbol{\phi}_{\chi',n}^\alpha \cdot \mathbf{b}_\chi^\alpha + \sum_{nm} \overline{\eta}_n^\alpha \overline{\eta}_m^\alpha \sum_\chi \tau_\chi P_\chi^0 \sum_{\chi'\chi''} W_{\chi\chi'} W_{\chi\chi''} \boldsymbol{\phi}_{\chi',n}^\alpha \cdot \boldsymbol{\phi}_{\chi'',m}^\alpha \\
&= \langle \mathbf{b}_\chi^\alpha \cdot \mathbf{b}_\chi^\alpha \tau_\chi \rangle_\chi + 2\sum_n \overline{\eta}_n^\alpha \langle \mathbf{b}_\chi^\alpha \cdot \sum_{\chi'} \Gamma_{\chi\chi'}\boldsymbol{\phi}_{\chi',n}^\alpha \rangle_\chi - 2\sum_n \overline{\eta}_n^\alpha \overline{b}_n^\alpha \\
&\quad + \sum_{nm} \overline{\eta}_n^\alpha \overline{\eta}_m^\alpha \sum_{\chi\chi'\chi''} P_{\chi'}^0 W_{\chi'\chi} \tau_\chi W_{\chi\chi''} \boldsymbol{\phi}_{\chi',n}^\alpha \cdot \boldsymbol{\phi}_{\chi'',m}^\alpha \\
&= \langle \mathbf{b}_\chi^\alpha \cdot \mathbf{b}_\chi^\alpha \tau_\chi \rangle_\chi + 2\sum_n \overline{\eta}_n^\alpha \langle \mathbf{b}_\chi^\alpha \cdot \sum_{\chi'} \Gamma_{\chi\chi'}\boldsymbol{\phi}_{\chi',n}^\alpha \rangle_\chi - 2\sum_n \overline{\eta}_n^\alpha \overline{b}_n^\alpha \\
&\quad + \sum_{nm} \overline{\eta}_n^\alpha \overline{\eta}_m^\alpha \left\{ \sum_{\chi\chi'\chi''} P_\chi^0 W_{\chi\chi'} \Gamma_{\chi'\chi''} \boldsymbol{\phi}_{\chi,n}^\alpha \cdot \boldsymbol{\phi}_{\chi'',m}^\alpha - \sum_{\chi\chi'} P_\chi^0 W_{\chi\chi'} \boldsymbol{\phi}_{\chi,n}^\alpha \cdot \boldsymbol{\phi}_{\chi',m}^\alpha \right\} \\
&= \langle \mathbf{b}_\chi^\alpha \cdot \mathbf{b}_\chi^\alpha \tau_\chi \rangle_\chi + 2\sum_n \overline{\eta}_n^\alpha \langle \mathbf{b}_\chi^\alpha \cdot \sum_{\chi'} \Gamma_{\chi\chi'}\boldsymbol{\phi}_{\chi',n}^\alpha \rangle_\chi - 2\sum_n \overline{\eta}_n^\alpha \overline{b}_n^\alpha + \sum_{nm} \overline{\eta}_n^\alpha \overline{\eta}_m^\alpha \left\{ \sum_{\chi\chi'} P_\chi^0 \tau_\chi^{-1}(\Gamma^2)_{\chi\chi'} \boldsymbol{\phi}_{\chi,n}^\alpha \cdot \boldsymbol{\phi}_{\chi',m}^\alpha \right. \\
&\quad \left. - 2\sum_{\chi\chi'} P_\chi^0 \tau_\chi^{-1} \Gamma_{\chi\chi'} \boldsymbol{\phi}_{\chi,n}^\alpha \cdot \boldsymbol{\phi}_{\chi',m}^\alpha + \sum_\chi P_\chi^0 \tau_\chi^{-1} \boldsymbol{\phi}_{\chi,n}^\alpha \cdot \boldsymbol{\phi}_{\chi,m}^\alpha \right\} \\
&= \langle \mathbf{b}_\chi^\alpha \cdot \mathbf{b}_\chi^\alpha \tau_\chi \rangle_\chi + 2\sum_n \overline{\eta}_n^\alpha \langle \mathbf{b}_\chi^\alpha \cdot \sum_{\chi'} \Gamma_{\chi\chi'}\boldsymbol{\phi}_{\chi',n}^\alpha \rangle_\chi - 2\sum_n \overline{\eta}_n^\alpha \overline{b}_n^\alpha - 2\sum_{nm} \overline{\eta}_n^\alpha \overline{\eta}_m^\alpha \overline{W}_{nm}^\alpha \\
&\quad + \sum_{nm} \overline{\eta}_n^\alpha \overline{\eta}_m^\alpha \left\{ \langle \sum_{\chi'} \tau_\chi^{-1}(\Gamma^2)_{\chi\chi'} \boldsymbol{\phi}_{\chi,n}^\alpha \cdot \boldsymbol{\phi}_{\chi',m}^\alpha \rangle_\chi - \langle \tau_\chi^{-1} \boldsymbol{\phi}_{\chi,n}^\alpha \cdot \boldsymbol{\phi}_{\chi',m}^\alpha \rangle_\chi \right\}
\end{aligned} \tag{S26}$$

Then, as $\overline{\eta}_n^\alpha = -\sum_m \overline{G}_{nm}^\alpha \overline{b}_m^\alpha$, we have

$$\begin{aligned}
R^\alpha &= \langle \mathbf{b}_\chi^\alpha \cdot \mathbf{b}_\chi^\alpha \tau_\chi \rangle_\chi - 2\sum_{nm} \langle \mathbf{b}_\chi^\alpha \cdot \sum_{\chi'} \Gamma_{\chi\chi'}\boldsymbol{\phi}_{\chi',n}^\alpha \rangle_\chi \overline{G}_{nm}^\alpha \overline{b}_m^\alpha \\
&\quad + \sum_{nn'mm'} \overline{b}_n^\alpha \overline{G}_{n'n}^\alpha \langle \tau_\chi^{-1}\boldsymbol{\phi}_{\chi,n'}^\alpha \cdot \left(-\boldsymbol{\phi}_{\chi,m'}^\alpha + \sum_{\chi'}(\Gamma^2)_{\chi\chi'}\boldsymbol{\phi}_{\chi',m'}^\alpha\right) \rangle_\chi \overline{G}_{m'm}^\alpha \overline{b}_m^\alpha.
\end{aligned} \tag{S27}$$

We can also identify a fractional residual bias as the ratio $R^\alpha$ to the first term in Eqn. S27.

## S5. BIAS BASIS APPROXIMATION METHOD FOR TRANSPORT COEFFICIENTS

A particularly simple chemisty-independent linear basis approximation is to use the bias vectors in each state, so that the index $n$ is replaced with the chemical index $\alpha$; that is, $\boldsymbol{\phi}_{\chi,\alpha} = \mathbf{b}_\chi^\alpha$. A related

9

choice is a scaled bias vector basis, $\boldsymbol{\phi}'_{\chi,\alpha} = \tau_\chi \mathbf{b}^\alpha_\chi$; it is also possible to simultaneously use both basis functions together, though that is not derived here. For the bias basis choice, we have the matrices and vectors

$$\underline{\overline{W}}_{\alpha\beta} = \left\langle \sum_{\chi'} W_{\chi\chi'} \mathbf{b}^\alpha_\chi \otimes \mathbf{b}^\beta_{\chi'} \right\rangle_\chi = \underline{\overline{W}}_{\beta\alpha}, \quad \overline{b}^\alpha_\beta = \left\langle \mathbf{b}^\alpha_\chi \cdot \mathbf{b}^\beta_\chi \right\rangle_\chi = \overline{b}^\beta_\alpha$$
$$\overline{W}_{\alpha\beta} = \operatorname{Tr}\underline{\overline{W}}_{\alpha\beta} = \left\langle \sum_{\chi'} W_{\chi\chi'} \mathbf{b}^\alpha_\chi \cdot \mathbf{b}^\beta_{\chi'} \right\rangle_\chi = \overline{W}_{\beta\alpha}$$
(S28)

or for the scaled bias basis choice, we have

$$\underline{\overline{W}}'_{\alpha\beta} = \left\langle \sum_{\chi'} W_{\chi\chi'} \tau_\chi \tau_{\chi'} \mathbf{b}^\alpha_\chi \otimes \mathbf{b}^\beta_{\chi'} \right\rangle_\chi = \underline{\overline{W}}'_{\beta\alpha}, \quad \overline{b}'^\alpha_\beta = \left\langle \tau_\chi \mathbf{b}^\alpha_\chi \cdot \mathbf{b}^\beta_\chi \right\rangle_\chi = \overline{b}'^\beta_\alpha$$
$$\overline{W}'_{\alpha\beta} = \operatorname{Tr}\underline{\overline{W}}'_{\alpha\beta} = \left\langle \sum_{\chi'} W_{\chi\chi'} \tau_\chi \tau_{\chi'} \mathbf{b}^\alpha_\chi \cdot \mathbf{b}^\beta_{\chi'} \right\rangle_\chi = \overline{W}'_{\beta\alpha}$$
(S29)

With these basis choices, the approximate transport coefficients are

$$k_B T V_0 \underline{L}^{(\alpha\beta)}_{BB} = \frac{1}{2}\left\langle \sum_{\chi'} W_{\chi\chi'} \delta\mathbf{x}^\alpha_{\chi\chi'} \otimes \delta\mathbf{x}^\beta_{\chi\chi'} \right\rangle_\chi + \sum_{\alpha'\beta'} \left\langle \mathbf{b}^{\alpha'}_\chi \cdot \mathbf{b}^\alpha_\chi \right\rangle_\chi \overline{G}_{\alpha'\beta'} \left\langle \mathbf{b}^{\beta'}_\chi \otimes \mathbf{b}^\beta_\chi \right\rangle_\chi$$
$$+ \sum_{\alpha'\beta'} \left\langle \mathbf{b}^\alpha_\chi \otimes \mathbf{b}^{\alpha'}_\chi \right\rangle_\chi \overline{G}_{\alpha'\beta'} \left\langle \mathbf{b}^{\beta'}_\chi \cdot \mathbf{b}^\beta_\chi \right\rangle_\chi - \sum_{\alpha'\alpha''\beta'\beta''} \left\langle \mathbf{b}^\alpha_\chi \cdot \mathbf{b}^{\alpha'}_\chi \right\rangle_\chi \overline{G}_{\alpha'\alpha''} \underline{\overline{W}}_{\alpha''\beta''} \overline{G}_{\beta''\beta'} \left\langle \mathbf{b}^{\beta'}_\chi \cdot \mathbf{b}^\beta_\chi \right\rangle_\chi$$
(S30)

and

$$k_B T V_0 \underline{L}^{(\alpha\beta)}_{SBB} = \frac{1}{2}\left\langle \sum_{\chi'} W_{\chi\chi'} \delta\mathbf{x}^\alpha_{\chi\chi'} \otimes \delta\mathbf{x}^\beta_{\chi\chi'} \right\rangle_\chi + \sum_{\alpha'\beta'} \left\langle \tau_\chi \mathbf{b}^{\alpha'}_\chi \cdot \mathbf{b}^\alpha_\chi \right\rangle_\chi \overline{G}'_{\alpha'\beta'} \left\langle \tau_\chi \mathbf{b}^{\beta'}_\chi \otimes \mathbf{b}^\beta_\chi \right\rangle_\chi$$
$$+ \sum_{\alpha'\beta'} \left\langle \tau_\chi \mathbf{b}^\alpha_\chi \otimes \mathbf{b}^{\alpha'}_\chi \right\rangle_\chi \overline{G}'_{\alpha'\beta'} \left\langle \tau_\chi \mathbf{b}^{\beta'}_\chi \cdot \mathbf{b}^\beta_\chi \right\rangle_\chi$$
$$- \sum_{\alpha'\alpha''\beta'\beta''} \left\langle \tau_\chi \mathbf{b}^\alpha_\chi \cdot \mathbf{b}^{\alpha'}_\chi \right\rangle_\chi \overline{G}'_{\alpha'\alpha''} \underline{\overline{W}}'_{\alpha''\beta''} \overline{G}'_{\beta''\beta'} \left\langle \tau_\chi \mathbf{b}^{\beta'}_\chi \cdot \mathbf{b}^\beta_\chi \right\rangle_\chi$$
(S31)

The primary advantage of these approximations is that they are easily expressible for non-lattice systems, and the matrices are the same dimensionality as the number of unique chemical species. The disadvantage is that the relaxation is short-ranged, which may be inappropriate for problems with complex energy landscapes.

## S6. AVERAGED GREEN FUNCTION APPROXIMATION FOR CUBIC RANDOM BINARY ALLOY

*a. Model.* We consider a model random binary system (chemistry A and B) on a cubic lattice, with a dilute vacancy concentration. We will identify our states $\chi = (n, \mathbf{x}_v)$ with the

position of a single vacancy, $\mathbf{x}_v$, and an integer vector $n_\mathbf{x}^\alpha$ that equals 0 or 1 if an atom of chemistry $\alpha$ is at the position $\mathbf{x}$. We have site exclusion so that $\sum_\alpha n_\mathbf{x}^\alpha = 1 - \delta_{\mathbf{x},\mathbf{x}_v}$ for all $\mathbf{x}$. For simplicity, we take the positions $\mathbf{x}$ to fall on a very large cubic lattice with Born von Karmann (periodic) boundary conditions, giving a total of $N \gg 1$ sites. We assume no interaction between A or B atoms, or the vacancy, so that $E_\chi$ is constant for all possible states. We have concentrations of A atoms equal to $c_A$ and of B atoms equal to $c_B = 1 - c_A$; we are interested in the limit $N \to \infty$, so the correction $c_v = N^{-1}$ in the concentrations due to the presence of a vacancy is negligible. Then,

$$P_\chi^0 = c_v \prod_\mathbf{x} (c_\alpha)^{n_\mathbf{x}^\alpha}$$

Finally, our rate model assumes that only the vacancy may exchange with neighboring sites, and that the rate depends only on the chemistry of the atom involved in the exchange: either $\nu_A$ or $\nu_B$.

b. *Basis functions.* We choose chemistry- and direction-independent basis functions, indexed by a lattice vector $\mathbf{x} \neq 0$ and chemistry $\beta \in \{A, B\}$,

$$\phi_{\chi,\beta \mathbf{x} i} = n_{\mathbf{x}+\mathbf{x}_v}^\beta \hat{e}_i,$$

that is, zero or one depending on whether an atom of chemistry $\beta$ is at a position $\mathbf{x}$ relative to the position of the vacancy in the state $\chi$. Then we have

$$\overline{W}_{\beta \mathbf{x}, \beta' \mathbf{x}'} = \left\langle \sum_{\chi'} W_{\chi \chi'} n_{\mathbf{x}+\mathbf{x}_v}^\beta n_{\mathbf{x}'+\mathbf{x}_v'}^{\beta'} \right\rangle_\chi, \quad \overline{\mathbf{b}}_{\beta \mathbf{x}}^\alpha = \left\langle \mathbf{b}_\chi^\alpha n_{\mathbf{x}+\mathbf{x}_v}^\beta \right\rangle_\chi \tag{S32}$$

Note that while $\beta, \beta' \in \{A, B\}$, $\alpha \in \{v, A, B\}$. This basis choice has the net effect of mapping the arbitrary solute concentration problem onto the dilute limit problem, where solute environments are averaged to produce rates and bias vectors. The only nonzero bias vectors $\overline{\mathbf{b}}_{\beta \mathbf{x}}^\alpha$ in Eqn. S32 are those $\mathbf{x}$ corresponding to a jump vector of a vacancy; similarly, $\overline{W}_{\alpha \mathbf{x}, \beta \mathbf{x}'}$ has a form similar to a vacancy transitioning through an averaged medium when $\mathbf{x}$ and $\mathbf{x}'$ do not correspond to a jump vector of a vacancy.

c. *Thermodynamic averages.* We define the average jump rate

$$\bar{\nu} := c_A \nu_A + c_B \nu_B$$

as it will appear throughout the problem. Let $\{\delta \mathbf{x}\}$ be the set of jump vectors for a vacancy on the lattice; in the case of a simple cubic lattice, these are $\{\pm a_0 \hat{x}, \pm a_0 \hat{y}, \pm a_0 \hat{z}\}$ for lattice constant $a_0$. For a given $\chi = (n, \mathbf{x}_v)$, the only $\chi' = (n', \mathbf{x}_v')$ with nonzero $W_{\chi \chi'}$ are those where

11

1. $\mathbf{x}'_v - \mathbf{x}_v \in \{\delta\mathbf{x}\}$

2. $n^\beta_\mathbf{x} = n'^\beta_\mathbf{x}$ for all $\mathbf{x} \notin \{\mathbf{x}_v, \mathbf{x}'_v\}, \beta \in \{A, B\}$

3. $n'^\beta_{\mathbf{x}_v} = n^\beta_{\mathbf{x}'_v}, \beta \in \{A, B\}$

That is: only two sites exchange, and atoms are conserved. We can then replace the sums over $\chi'$ instead with a sum over possible $\delta\mathbf{x}$, and simply construct both the $\mathbf{x}'_v = \mathbf{x} + \delta\mathbf{x}$ and $n'$ that satisfies the above conditions. When performing all averages, only the sites that are exchanging with a vacancy or have an explicit occupancy variable need to be treated.

d. *Unbiased contribution.* The unbiased contribution to the transport coefficients is

$$\frac{1}{2}\left\langle \sum_{\chi'} W_{\chi\chi'} \delta\mathbf{x}^\alpha_{\chi\chi'} \otimes \delta\mathbf{x}^\beta_{\chi\chi'} \right\rangle_\chi = \mathbf{1} c_v a_0^2 \begin{cases} \bar{\nu} & : \alpha = \beta = v \\ c_\alpha \nu_\alpha & : \alpha = \beta \in \{A, B\} \\ -c_\beta \nu_\beta & : \alpha = v, \beta \in \{A, B\} \text{ or } \beta = v, \alpha \in \{A, B\} \text{ or} \\ 0 & : \alpha \neq \beta \in \{A, B\} \end{cases} \quad (S33)$$

e. *Bias vectors.* Then, the averaged bias vectors are

$$\overline{\mathbf{b}}^\alpha_{\beta\mathbf{x}} = c_v \begin{cases} c_\beta(\nu_\beta - \bar{\nu})\mathbf{x} & : \mathbf{x} \in \{\delta\mathbf{x}\}, \alpha = v \\ -c_A c_B \nu_\alpha \mathbf{x} & : \mathbf{x} \in \{\delta\mathbf{x}\}, \alpha = \beta \in \{A, B\} \\ c_A c_B \nu_\alpha \mathbf{x} & : \mathbf{x} \in \{\delta\mathbf{x}\}, \alpha \neq \beta \in \{A, B\} \\ 0 & : \text{otherwise} \end{cases} \quad (S34)$$

Note that on a cubic lattice, non-zero bias is only possible when a vacancy sits between an A and B atom at $+\delta\mathbf{x}$ and $-\delta\mathbf{x}$, respectively. This configuration always has a probability factor of $c_A c_B$. The $\beta = \alpha$ and $\beta \neq \alpha$ cases identify whether the vector $x$ is pointing to the $\alpha$ atom in this A-v-B "complex," or its opposite; hence the sign change. For the case where $\alpha = v$, the same probability factor applies, while the rate is either $\nu_A - \nu_B$ for $\beta = A$ or $\nu_B - \nu_A$ for $\beta = B$. The combination

$$c_\beta(\nu_\beta - \bar{\nu}) = \begin{cases} c_A(\nu_A - c_A\nu_A - c_B\nu_B) = c_A c_B(\nu_A - \nu_B) & : \beta = A \\ c_B(\nu_B - c_A\nu_A - c_B\nu_B) = c_A c_B(\nu_B - \nu_A) & : \beta = B \end{cases} \quad (S35)$$

as expected. Moreover, $\sum_\alpha \overline{\mathbf{b}}^\alpha_{\beta\mathbf{x}} = 0$ for all $\beta\mathbf{x}$.

*f. Rate matrix.* To simplify the rate matrix problem, we first construct the "far field" form, when $\mathbf{x}, \mathbf{x}' \notin \{\boldsymbol{\delta x}\}$. Define the $\overline{W}^0$ matrix as

$$\overline{W}^0_{\beta\mathbf{x},\beta'\mathbf{x}'} := c_v \begin{cases} z(c_\beta c_{\beta'} - \delta_{\beta\beta'} c_\beta)\bar{\nu} & : \mathbf{x} = \mathbf{x}' \\ (\delta_{\beta\beta'} c_\beta - c_\beta c_{\beta'})\bar{\nu} & : \mathbf{x} - \mathbf{x}' \in \{\boldsymbol{\delta x}\} \\ 0 & : \text{otherwise} \end{cases} \quad (S36)$$

where $z$ is the coordination of our lattice (6 for simple cubic). In this case, the rate matrix allows $\mathbf{x}$ and $\mathbf{x}'$ to be $\mathbf{0}$; this does not correspond to a basis function in our problem, but rather, is an artificially added state to give $\overline{W}^0$ translational invariance. The probability factor—which comes from any solute-solute correlation—is $\delta_{\beta\beta'} c_\beta - c_\beta c_{\beta'}$ for the random alloy, and equals $c_A c_B$ for $\beta = \beta'$ and $-c_A c_B$ for $\beta \neq \beta'$. This means that $\overline{W}^0_{A\mathbf{x},A\mathbf{x}'} = \overline{W}^0_{B\mathbf{x},B\mathbf{x}'} = -\overline{W}^0_{A\mathbf{x},B\mathbf{x}'}$, and a similar symmetry holds for $\overline{W}$. Now, we write $\overline{W} = \overline{W}^0 + \overline{\delta W}$ where

$$\overline{\delta W}_{A\mathbf{x},A\mathbf{x}'} := c_v \begin{cases} zc_A c_B \bar{\nu} & : \mathbf{x} = \mathbf{x}' = \mathbf{0} \\ -c_A c_B \bar{\nu} & : \mathbf{x} = \mathbf{0}, \mathbf{x}' \in \{\boldsymbol{\delta x}\} \text{ or } \mathbf{x}' = \mathbf{0}, \mathbf{x} \in \{\boldsymbol{\delta x}\} \text{ or} \\ c_A c_B (\nu_A + \nu_B - \bar{\nu}) & : \mathbf{x} = -\mathbf{x}' \in \{\boldsymbol{\delta x}\} \\ -c_A c_B (\nu_A + \nu_B - 2\bar{\nu}) & : \mathbf{x} = \mathbf{x}' \in \{\boldsymbol{\delta x}\} \\ 0 & : \text{otherwise} \end{cases} \quad (S37)$$

while $\overline{\delta W}_{A\mathbf{x},A\mathbf{x}'} = \overline{\delta W}_{B\mathbf{x},B\mathbf{x}'} = -\overline{\delta W}_{A\mathbf{x},B\mathbf{x}'}$. The terms corresponding to $\mathbf{x} = \mathbf{0}$ or $\mathbf{x}' = \mathbf{0}$ are corrections that remove the "artificial" basis functions that were added previously. Meanwhile, the $\mathbf{x}' = -\mathbf{x}$ corresponds the sum of two contributions: an A atom at $\mathbf{x}$ relative to a vacancy is exchanged and arrives at $-\mathbf{x}$ relative to the vacancy, *and* the change in negative escape rate from a state that has an A atom at both $\pm\mathbf{x}$ which can exchange with the vacancy. Finally, the $\mathbf{x}' = \mathbf{x}$ term is the change in the negative escape rate due to the $\mathbf{x}' = -\mathbf{x}$ term and the $\mathbf{x}' = \mathbf{0}$ term.

*g. Green function solution.* This formulation gives the problem a similar form to the *dilute* concentration limit, as we are considering only the interaction between a single vacancy and a single site; albeit now environments are averaged. It is also worth noting that the symmetry of the $\overline{W}$ matrix with respect to $\beta, \beta'$ requires the use of pseudoinverse. This is because our basis choice, while symmetric, is also linearly dependent, as $n_\mathbf{x}^A = 1 - n_\mathbf{x}^B$ when $\mathbf{x} \neq \mathbf{x}_v$ for any $\chi = (n, \mathbf{x}_v)$. That minor complication only means that $\overline{G}_{\beta\mathbf{x},\beta'\mathbf{x}'}$ will share the same symmetry with respect to $\beta, \beta'$ that $\overline{W}$ does. We can solve for our Green function using a Dyson equation approach where

13

$(\overline{G})^{-1} = (\overline{G}^0)^{-1} + \overline{\delta W}$; the $\overline{G}^0$ function corresponds to the Green function for a bare vacancy jumping with a probability-rate product of $c_A c_B \bar{\nu}$. For a cubic lattice, including simple cubic, we only need to compute $\overline{G}^0$ for $\mathbf{x} - \mathbf{x}' \in \{\mathbf{0}, 2\delta\mathbf{x}\}$. In particular, due to symmetry, we need to know the difference $\overline{G}^0_{A\mathbf{0},A\mathbf{0}} - \overline{G}^0_{A\mathbf{0},A2\delta\mathbf{x}} = \gamma^{-1}(c_A c_B \bar{\nu})^{-1}$, where $\gamma$ is a crystal structure dependent constant, related to the dilute-limit tracer correlation factor $f$ by

$$\gamma = \frac{f+1}{f-1}$$

Since $0 < f < 1$, then $\gamma < -1$. Finally, we can solve for our transport coefficients

$$\begin{aligned}
\underline{L}^{(vv)} &= \mathbf{1} c_v a_0^2 \left[ \bar{\nu} - c_A c_B \frac{(\nu_A - \nu_B)^2}{\nu_A + \nu_B - \frac{3+\gamma}{2}\bar{\nu}} \right] \\
\underline{L}^{(AA)} &= \mathbf{1} c_v a_0^2 \left[ c_A \nu_A - c_A c_B \frac{\nu_A^2}{\nu_A + \nu_B - \frac{3+\gamma}{2}\bar{\nu}} \right] \\
\underline{L}^{(BB)} &= \mathbf{1} c_v a_0^2 \left[ c_B \nu_B - c_A c_B \frac{\nu_B^2}{\nu_A + \nu_B - \frac{3+\gamma}{2}\bar{\nu}} \right] \\
\underline{L}^{(vA)} &= \mathbf{1} c_v a_0^2 \left[ -c_A \nu_A + c_A c_B \frac{\nu_A(\nu_A - \nu_B)}{\nu_A + \nu_B - \frac{3+\gamma}{2}\bar{\nu}} \right] \\
\underline{L}^{(vB)} &= \mathbf{1} c_v a_0^2 \left[ -c_B \nu_B + c_A c_B \frac{\nu_B(\nu_B - \nu_A)}{\nu_A + \nu_B - \frac{3+\gamma}{2}\bar{\nu}} \right] \\
\underline{L}^{(AB)} &= \mathbf{1} c_v a_0^2 \left[ c_A c_B \frac{\nu_A \nu_B}{\nu_A + \nu_B - \frac{3+\gamma}{2}\bar{\nu}} \right]
\end{aligned} \quad (S38)$$

The correlation factor in the denominator can be written as

$$-\frac{3+\gamma}{2} = -\frac{2f-1}{1-f} = 1 - \frac{f}{1-f}$$

Note finally that our Onsager matrix is symmetric, and sums to zero along any column / row. This result reproduces the dilute limit result exactly, as expected.

## S7. RESIDUAL BIAS CORRECTION

Given our variational approach, we can also apply more than one approximation method in succession to build more accurate results. One example is to use the *residual bias vectors* from a linear basis approximation method as a new basis to derive a correction to the LBAM diffusivity results. If we use the scaled residual bias, then the second-order basis functions are $\boldsymbol{\psi}_{\chi,\alpha} := \tau_\chi \widetilde{\mathbf{b}}_\chi^\alpha$

where $\widetilde{\mathbf{b}}_\chi^\alpha$ is the bias remaining after the LBAM solution; and the relevant expressions in Eqn. S29 and Eqn. S31 can be derived directly, with reference to the $\overline{\eta}_n^\alpha$ expressions, as in Eqn. S27.

First, our transport coefficients are

$$k_\mathrm{B} T V_0 \underline{L}_{\mathrm{LBAM+RBC}}^{(\alpha\beta)} = k_\mathrm{B} T V_0 \underline{L}_{\mathrm{LBAM}}^{(\alpha\beta)} + \langle \boldsymbol{\psi}_{\chi,\alpha} \otimes \widetilde{\mathbf{b}}_\chi^\beta \rangle_\chi (\overline{W}_\mathrm{r}^\beta)^{-1} \langle \boldsymbol{\psi}_{\chi,\alpha} \cdot \widetilde{\mathbf{b}}_\chi^\alpha \rangle_\chi$$
$$+ \langle \widetilde{\mathbf{b}}_\chi^\alpha \otimes \boldsymbol{\psi}_{\chi,\beta} \rangle_\chi (\overline{W}_\mathrm{r}^\beta)^{-1} \langle \boldsymbol{\psi}_{\chi,\beta} \cdot \widetilde{\mathbf{b}}_\chi^\beta \rangle_\chi - \langle \boldsymbol{\psi}_{\chi,\alpha} \cdot \widetilde{\mathbf{b}}_\chi^\alpha \rangle_\chi (\overline{W}_\mathrm{r}^\alpha)^{-1} \overline{W}_\mathrm{r}^{\alpha\beta} (\overline{W}_\mathrm{r}^\beta)^{-1} \langle \boldsymbol{\psi}_{\chi,\beta} \cdot \widetilde{\mathbf{b}}_\chi^\beta \rangle_\chi$$
(S39)

where the corresponding terms are written in terms of the $\Gamma$ matrix (c.f. Section S3),

$$\langle \boldsymbol{\psi}_{\chi,\alpha} \otimes \widetilde{\mathbf{b}}_\chi^\beta \rangle_\chi = \langle \tau_\chi \widetilde{\mathbf{b}}_\chi^\alpha \otimes \widetilde{\mathbf{b}}_\chi^\beta \rangle_\chi$$
$$= \langle \tau_\chi \mathbf{b}_\chi^\alpha \otimes \mathbf{b}_\chi^\beta \rangle_\chi - \sum_n \langle \mathbf{b}_\chi^\alpha \otimes \boldsymbol{\phi}_{\chi,n}^\beta \rangle_\chi \overline{\eta}_n^\beta + \sum_n \Big\langle \sum_{\chi'} \Gamma_{\chi\chi'} \mathbf{b}_\chi^\alpha \otimes \boldsymbol{\phi}_{\chi',n}^\beta \Big\rangle_\chi \overline{\eta}_n^\beta$$
$$- \sum_n \overline{\eta}_n^\alpha \langle \boldsymbol{\phi}_{\chi,n}^\alpha \otimes \mathbf{b}_\chi^\beta \rangle_\chi + \sum_n \overline{\eta}_n^\alpha \Big\langle \sum_{\chi'} \Gamma_{\chi\chi'} \boldsymbol{\phi}_{\chi',n}^\alpha \otimes \mathbf{b}_\chi^\beta \Big\rangle_\chi$$
$$+ \sum_{nn'} \overline{\eta}_n^\alpha \Big\langle \sum_{\chi'} \tau_\chi^{-1} [\Gamma^2 - \mathbf{1}]_{\chi\chi'} \boldsymbol{\phi}_{\chi,n}^\alpha \otimes \boldsymbol{\phi}_{\chi',n'}^\beta \Big\rangle_\chi \overline{\eta}_{n'}^\beta - 2 \sum_{nn'} \overline{\eta}_n^\alpha \overline{W}_\mathrm{r}^{\alpha\beta} \overline{\eta}_{n'}^\beta$$
(S40)

and

$$\langle \boldsymbol{\psi}_{\chi,\alpha} \cdot \widetilde{\mathbf{b}}_\chi^\alpha \rangle_\chi = \langle \tau_\chi \mathbf{b}_\chi^\alpha \cdot \mathbf{b}_\chi^\alpha \rangle_\chi + 2 \sum_n \Big\langle \sum_{\chi'} \Gamma_{\chi\chi'} \mathbf{b}_\chi^\alpha \cdot \boldsymbol{\phi}_{\chi',n}^\alpha \Big\rangle_\chi \overline{\eta}_n^\alpha$$
$$+ \sum_{nn'} \overline{\eta}_n^\alpha \Big\langle \sum_{\chi'} \tau_\chi^{-1} [\Gamma^2 - \mathbf{1}]_{\chi\chi'} \boldsymbol{\phi}_{\chi,n}^\alpha \cdot \boldsymbol{\phi}_{\chi',n'}^\alpha \Big\rangle_\chi \overline{\eta}_{n'}^\alpha$$
(S41)

and

$$\overline{W}_\mathrm{r}^{\alpha\beta} = \Big\langle \sum_{\chi'} W_{\chi\chi'} \tau_\chi \widetilde{\mathbf{b}}_\chi^\alpha \otimes \tau_{\chi'} \widetilde{\mathbf{b}}_{\chi'}^\beta \Big\rangle_\chi$$
$$= \Big\langle \sum_{\chi'} \Gamma_{\chi\chi'} \mathbf{b}_\chi^\alpha \otimes \mathbf{b}_{\chi'}^\beta \tau_{\chi'} \Big\rangle_\chi - \langle \tau_\chi \mathbf{b}_\chi^\alpha \otimes \mathbf{b}_\chi^\beta \rangle_\chi$$
$$+ \sum_n \overline{\eta}_n^\alpha \Big\langle \sum_{\chi'} [\Gamma^2 - 2\Gamma + \mathbf{1}]_{\chi\chi'} \boldsymbol{\phi}_{\chi',n}^\alpha \otimes \mathbf{b}_\chi^\beta \Big\rangle_\chi + \sum_n \Big\langle \sum_{\chi'} [\Gamma^2 - 2\Gamma + \mathbf{1}]_{\chi\chi'} \mathbf{b}_\chi^\alpha \otimes \boldsymbol{\phi}_{\chi',n}^\beta \Big\rangle_\chi \overline{\eta}_n^\beta$$
$$+ \sum_{nn'} \overline{\eta}_n^\alpha \Big\langle \sum_{\chi'} \tau_\chi^{-1} [\Gamma^3 - 3\Gamma^2 + 3\Gamma - \mathbf{1}]_{\chi\chi'} \boldsymbol{\phi}_{\chi,n}^\alpha \otimes \boldsymbol{\phi}_{\chi',n'}^\beta \Big\rangle_\chi \overline{\eta}_{n'}^\beta$$
(S42)

and

$$\overline{W}_\mathrm{r}^\alpha = \Big\langle \sum_{\chi'} \Gamma_{\chi\chi'} \mathbf{b}_\chi^\alpha \cdot \mathbf{b}_{\chi'}^\alpha \tau_{\chi'} \Big\rangle_\chi - \langle \tau_\chi \mathbf{b}_\chi^\alpha \cdot \mathbf{b}_\chi^\alpha \rangle_\chi + 2 \sum_n \Big\langle \sum_{\chi'} [\Gamma^2 - 2\Gamma + \mathbf{1}]_{\chi\chi'} \mathbf{b}_\chi^\alpha \cdot \boldsymbol{\phi}_{\chi',n}^\alpha \Big\rangle_\chi \overline{\eta}_n^\alpha$$
$$+ \sum_{nn'} \overline{\eta}_n^\alpha \Big\langle \sum_{\chi'} \tau_\chi^{-1} [\Gamma^3 - 3\Gamma^2 + 3\Gamma - \mathbf{1}]_{\chi\chi'} \boldsymbol{\phi}_{\chi,n}^\alpha \cdot \boldsymbol{\phi}_{\chi',n'}^\alpha \Big\rangle_\chi \overline{\eta}_{n'}^\alpha$$
(S43)

15

In the case of a chemistry- and direction-independent basis $\phi_{\chi,n}$, some additional simplification are possible, as before. In that case,

$$\langle \psi_{\chi,\alpha} \otimes \widetilde{\mathbf{b}}_\chi^\beta \rangle_\chi = \langle \tau_\chi \mathbf{b}_\chi^\alpha \otimes \mathbf{b}_\chi^\beta \rangle_\chi + \sum_n \Big\langle \sum_{\chi'} \Gamma_{\chi\chi'} \mathbf{b}_\chi^\alpha \phi_{\chi',n} \Big\rangle_\chi \otimes \overline{\eta}_n^\beta + \sum_n \overline{\eta}_n^\alpha \otimes \Big\langle \sum_{\chi'} \Gamma_{\chi\chi'} \mathbf{b}_\chi^\beta \phi_{\chi',n} \Big\rangle_\chi$$

$$+ \sum_{nn'} \overline{\eta}_n^\alpha \Big\langle \sum_{\chi'} \tau_\chi^{-1} [\Gamma^2 - \mathbf{1}]_{\chi\chi'} \phi_{\chi,n} \phi_{\chi',n'} \Big\rangle_\chi \otimes \overline{\eta}_{n'}^\beta$$

$$\langle \psi_{\chi,\alpha} \cdot \widetilde{\mathbf{b}}_\chi^\alpha \rangle_\chi = \langle \tau_\chi \mathbf{b}_\chi^\alpha \cdot \mathbf{b}_\chi^\alpha \rangle_\chi + 2\sum_n \Big\langle \sum_{\chi'} \Gamma_{\chi\chi'} \mathbf{b}_\chi^\alpha \phi_{\chi',n} \Big\rangle_\chi \cdot \overline{\eta}_n^\alpha$$

$$+ \sum_{nn'} \overline{\eta}_n^\alpha \Big\langle \sum_{\chi'} \tau_\chi^{-1} [\Gamma^2 - \mathbf{1}]_{\chi\chi'} \phi_{\chi,n} \phi_{\chi',n'} \Big\rangle_\chi \cdot \overline{\eta}_{n'}^\alpha$$

$$\underline{\overline{W}}_{\mathrm{r}}^{\alpha\beta} = \Big\langle \sum_{\chi'} \Gamma_{\chi\chi'} \mathbf{b}_\chi^\alpha \otimes \mathbf{b}_{\chi'}^\beta \tau_{\chi'} \Big\rangle_\chi - \langle \tau_\chi \mathbf{b}_\chi^\alpha \otimes \mathbf{b}_\chi^\beta \rangle_\chi + \frac{1}{2} \sum_n (\overline{\eta}_n^\alpha \otimes \overline{\mathbf{b}}_n^\beta + \overline{\mathbf{b}}_n^\alpha \otimes \overline{\eta}_n^\beta)$$

$$+ \sum_n \overline{\eta}_n^\alpha \otimes \Big\langle \sum_{\chi'} [\Gamma(\Gamma - 2)]_{\chi\chi'} \mathbf{b}_\chi^\beta \phi_{\chi',n} \Big\rangle_\chi + \sum_n \Big\langle \sum_{\chi'} [\Gamma(\Gamma - 2)]_{\chi\chi'} \mathbf{b}_\chi^\alpha \phi_{\chi',n} \Big\rangle_\chi \otimes \overline{\eta}_n^\beta$$

$$+ \sum_{nn'} \overline{\eta}_n^\alpha \otimes \Big\langle \sum_{\chi'} [W\Gamma(\Gamma - 2)]_{\chi\chi'} \phi_{\chi,n} \phi_{\chi',n'} \Big\rangle_\chi \overline{\eta}_{n'}^\beta$$

$$\overline{W}_{\mathrm{r}}^\alpha = \Big\langle \sum_{\chi'} \Gamma_{\chi\chi'} \mathbf{b}_\chi^\alpha \cdot \mathbf{b}_{\chi'}^\alpha \tau_{\chi'} \Big\rangle_\chi - \langle \tau_\chi \mathbf{b}_\chi^\alpha \cdot \mathbf{b}_\chi^\alpha \rangle_\chi + \sum_n \overline{\eta}_n^\alpha \cdot \overline{\mathbf{b}}_n^\alpha$$

$$+ 2\sum_n \Big\langle \sum_{\chi'} [\Gamma(\Gamma - 2)]_{\chi\chi'} \mathbf{b}_\chi^\alpha \phi_{\chi',n} \Big\rangle_\chi \cdot \overline{\eta}_n^\alpha$$

$$+ \sum_{nn'} \overline{\eta}_n^\alpha \cdot \Big\langle \sum_{\chi'} [W\Gamma(\Gamma - 2)]_{\chi\chi'} \phi_{\chi,n} \phi_{\chi',n'} \Big\rangle_\chi \overline{\eta}_{n'}^\alpha$$

(S44)

Finally, for our square lattice, we can numerically evaluate the expressions as polynomials in $c_\mathrm{B}$ for the case where $\nu_\mathrm{B} = 0$ to compute an analytic expression for the diffusivity of solvent that is a correction to the averaged Green function result above. This gives, for the square lattice,

$$D_\mathrm{A} = \frac{1 - c_\mathrm{B}}{1 + (3 - \pi)c_\mathrm{B}} + \Big\{ -0.01272990905 c_\mathrm{B} + 4.529059154 c_\mathrm{B}^2 - 399.7080744 c_\mathrm{B}^3$$
$$- 561.6483202 c_\mathrm{B}^4 + 665.0100411 c_\mathrm{B}^5 + 622.9427624 c_\mathrm{B}^6 - 379.2388949 c_\mathrm{B}^7 + 48.12615674 c_\mathrm{B}^8 \Big\} \Big/$$
$$\Big\{ 1 + 361.2297602 c_\mathrm{B} + 590.7833342 c_\mathrm{B}^2 + 222.4121227 c_\mathrm{B}^3 + 307.7589952 c_\mathrm{B}^4$$
$$+ 208.3266238 c_\mathrm{B}^5 - 52.05560275 c_\mathrm{B}^6 - 24.0423294 c_\mathrm{B}^7 - 1.884593043 c_\mathrm{B}^8 \Big\}$$

(S45)

where the first term is the analytic two-body Green function result.

16


\end{document}